\documentclass[preprint]{aastex}
\usepackage{epsfig,lscape}

\def\gap{\lower.5ex\hbox{$\; \buildrel > \over \sim \;$}}

\def\cm-2{cm$^{-2}$}

\def\xmm{{XMM-Newton }}

\def\n253{\object{NGC~253}}
\def\m33{\object{M~33}}
\def\lap{\lower.5ex\hbox{$\; \buildrel < \over \sim \;$}}
\def\mx7{\object{M~33~X$-$7}}
\def\x7{\hbox{X$-$7}}

\begin{document}

\title{A Spectroscopic Search for High Mass X-ray Binaries in M31}

   \author{B. F. ~Williams\altaffilmark{1}, D.~Hatzidimitriou\altaffilmark{2}, J. ~Green\altaffilmark{3},
   G.~Vasilopoulos\altaffilmark{4}, R. ~Covarrubias\altaffilmark{5}, W.N. ~Pietsch\altaffilmark{4}, H.
~Stiele\altaffilmark{6}, F. ~Haberl\altaffilmark{4}, P. ~Bonfini\altaffilmark{7} }

\altaffiltext{1}{Department of Astronomy, Box 351580, University of
Washington, Seattle, WA 98195}
\altaffiltext{2}{University of
Athens, Faculty of Physics, Department of Astrophysics, Astronomy
and Mechanics, Panepistimiopolis, GR15784 Zografos, Athens, Greece}
\altaffiltext{3}{Center for Astrophysics, 60 Garden Street,
Cambridge, MA 06540}
 \altaffiltext{4}{Max-Planck-Institut f\"ur
extraterrestrische Physik, 85741 Garching, Germany}
\altaffiltext{5}{National Optical Astronomy Observatories, Colina El
Pino, Casilla 603 La Serena, Chile}
 \altaffiltext{6}{Shanghai
Astronomical Observatory, 80 Nandan Road, Shanghai 200030 China}
 \altaffiltext{7}{Centre for Astrophysics and Supercomputing, Mail number H30
Swinburne University of Technology, PO Box 218, Hawthorn, Victoria 3122, Australia}


\begin{abstract}

We present new optical spectroscopy of 20 candidate counterparts of 17
X-ray sources in the direction of the M31 disc.  By comparing the
X-ray catalogue from the XMM-Newton survey of M31 with star catalogues
from the Local Group Galaxy Survey, we chose counterpart candidates
based on optical colour and X-ray hardness.  We have discovered 17
counterpart candidates with spectra containing stellar features.
Eight of these are early-type stars of O or B type in M31, with hard
X-ray spectra, making them good HMXB candidates.  Three of these eight
exhibit emission lines, which we consider to be the strongest HMXB
candidates.  In addition, our spectra reveal two likely Galactic
cataclysmic variables, one foreground M star, two probable LMXBs
related to M31 globular clusters, one emission line region with an
embedded Wolf-Rayet star, and one newly-discovered supernova remnant.
Finally, two of the sources have stellar spectra with no features
indicative of association with an X-ray source.

\end{abstract}

\section{Introduction}

One puzzling mystery surrounding our neighboring spiral galaxy M31 is
that while the Milky Way, which is comparable in mass and morphology,
has 130 known high-mass X-ray binaries (HMXBs), M31 has none.
Galactic high-mass X-ray binaries (HMXBs) typically contain stars of
spectral type O8--B3 \citep{liu2000}. These stars have $-5< M_V
<-2$, or $19.7<V <22.7$ when scaled to the distance
of \citep[780 kpc;][]{holland1996,mcconnachie2005}
 and reddening \citep[$A_V=0.3$][]{dalcanton2012}
of M31.  We have performed a spectral survey of bright blue optical
counterpart candidates of catalogued X-ray sources in an attempt to
discover the  HMXB population of M31, and to begin measuring
its characteristics.

Being young objects, HMXBs are associated with star formation. In M31,
star formation is mostly associated with the disc. An investigation of
the star formation rate of the disc has been conducted by
\citet{williams2003}, who finds that the mean SFR over the last 60 Myr
for 1.4 deg$^2$ of the M31 disc is $0.63{\pm}0.07~{\rm
M}_{\odot}$~yr$^{-1}$ (in the range 0.1-100 $M_{\odot}$) with no
drastic changes.  We used the calibration of \citet{Grimm2003} to
calculate the expected number of HMXBs (see comment in
\citealp{ShtykovskiyandGilfanov2005}, regarding the normalization) to
be $\simeq$2 HMXBs brighter than $10^{38}$ erg~s$^{-1}$, and
$\simeq$32 sources brighter than $10^{36}$ erg~s$^{-1}$.

Alternatively, recent work on the luminosity function of HMXBs in the
Small Magellanic Cloud has resulted in a different shape than for
larger galaxies \citep{sturm2013}.  While this shape may not be
relevant for the larger galaxy M31, it is more sharply cut off at
luminosities $>$10$^{36}$ erg s$^{-1}$.  Scaling the SMC HMXB XLF by a
factor of 6 to account for the higher star formation rate in M31, we
would expect $\simeq$40 HMXBs with a luminosity above 10$^{36}$
erg~s$^{-1}$, but none brighter than 10$^{37}$ erg~s$^{-1}$.

The estimates are reasonable, given that the Galaxy has $\simeq$20 HMXBs
with X-ray luminosities $>$10$^{36}$ erg s$^{-1}$ \citep{grimm2002},
which could be detected at the distance of M31.  However, HMXBs are
difficult to find at that distance, as evidenced by the fact that M33,
at similar distance and star formation rate but with lower star
density and dust content, has just 3 known HMXBs
\citep{long2002,pietsch2006,pietsch2009}, one of which has been
confirmed with radial velocities \citep{orosz2007}.

M31 has been well-observed and catalogued in the X-ray and optical
\citep{williams2004a,pietsch2005,stiele2011,massey2006,dalcanton2012},
so one would expect that HMXBs would have been discovered and
confirmed by now, as is the case with M33; however, there are several
possible reasons why we have not yet found any confirmed HMXBs in M31.
The most likely is that they are more heavily crowded by other stars
and more heavily reddened than in M33.  Another possibility is that
the relatively low star formation rate, and high metallicity
\citep[solar--supersolar;][]{zaritsky1994} may hamper HMXB production
\citep{dray2006}.  Another possibility is that many of the HMXBs in
M31 are too faint to be detected by our currently-available surveys.

While deep {\it Chandra} observations of M31 are mainly limited to the
central portions of the galaxy
\citep{kaaret2002,williams2006,voss2007,garcia2010,barnard2013,hofmann2013},
where spectroscopic follow-up is very difficult, \xmm has surveyed the
entire galaxy \citep{stiele2011}.  \xmm EPIC observations taken
between June 2006 and February 2008, together with archival
observations \citep{pietsch2005} from June 2000 to July 2004 yielded a
total of 1948 X-ray sources (0.2-12.0 keV) covering (for the first
time) the entire D25 ellipse of M 31, down to a limiting luminosity of
$\simeq 10^{35}$ erg~s$^{-1}$ in the 0.2-4.5 keV band
\citep{stiele2011}. These sources include (i) sources within M31,
i.e. X-ray binaries, supernova remnants (SNRs) and supersoft sources, (ii)
foreground (Galactic) stars and (iii) background objects, i.e.  mostly
active galactic nuclei, some normal galaxies and a few clusters of
galaxies.  Hardness ratios, X-ray variability and correlations with
catalogues in other wavelengths (optical, radio) have been used in an
attempt to classify the various X-ray sources.  However, a high
percentage (65\%) of the sources can still only be classified as
$<$hard$>$ sources, i.e. it is not possible to determine whether these
sources are X-ray binaries or Crab-like SNRs in M31, or
background X-ray sources (mainly AGNs) not associated with M31.

Here, we discuss spectroscopic results regarding the X-ray binary
population revealed by these X-ray data and follow-up optical
spectroscopic observations of counterpart candidates found within
these data sets, including 8 bright, early-type stars that are
probable HMXBs.  Two of these 8 have spectra with emission lines,
making them good HMXB candidates.  Another one of these 8 has a
spectrum similar to Be stars, making it a strong Be-HMXB candidate.
Many of the other spectra obtained were newly-discovered AGN behind
M31 which will be discussed in another paper (Green et al., in prep.).
Section 2 describes how our counterpart candidates were selected, the
data acquisition and reduction process, and the classification of the
optical spectra.  Section 3 details our results, providing
descriptions of each HMXB candidate individually, and Section 4
summarizes our findings.

\section{Data}

\subsection{Counterpart Selection}

Optical counterparts of the X-ray sources were sought within the
optical catalogue of M31 from the Local Group Galaxy Survey
\cite[LGGS][]{massey2006}, which provides stellar photometry covering
the entire optical disc of M31. The photometric accuracy is 1\% down
to 21st magnitude, and the positional accuracy is 0.25$''$.  We
cross-correlated the \citet{pietsch2005} and \citet{massey2006}
catalogues, using a $3\sigma$ (combined X-ray and optical positional
uncertainty) search radius.  Typical search radii were $\simeq$4$''$,
and the local densities of bright blue stars in the \citet{massey2006}
catalogue within 0.1$^{\circ}$ of our spectroscopy targets was $\leq$6
arcmin$^{-2}$, making the chance superposition rate $\leq$8\%.

The optical counterparts of HMXBs are expected to be Be stars or
supergiants. Using the V-band magnitude and colour limits for Be stars
from \citep{sabogal2005}, and taking into account the distance modulus
and interstellar absorption of M31,  we have estimated blue supergiants
and Be stars in M31 to have apparent magnitudes in the range V$=17-25$
and $-0.4<B-V<0.8$, with the bright end (V$<$19.5) being supergiants
and the faint end V($>$19.5) being Be stars. Among the optical
counterparts of X-ray sources mentioned above, down to the limiting
magnitude of 20.5 (where the \citealp{massey2006} is highly complete),
there are 91 objects within the expected colour range.  However, with
our available telescope and instrument (described below), we were able
to obtain spectra for $\sim$6 objects per season.  Since about half of
the observed sources turned out not to be stellar, our final stellar
sample for this work was 20 spectra of 17 objects in the 7 season
campaign.

\subsection{Optical observations}\label{observations}

The optical data presented here were acquired with the 3.5m
telescope at the Apache Point Observatory (APO), located in the
Sacramento Mountains in Sunspot, New Mexico (USA). Observations were
carried out during fall of 2006, 2008 and 2011,  during summer and
fall of 2007, 2009,  fall of 2011 and summer of 2012.

The telescope was equipped with the Dual Imaging Spectrograph (DIS), a
medium dispersion spectrograph with separate collimators for the red
and blue part of the spectrum and two 2048x1028 E2V CCD cameras, with
the transition wavelength around 5350\AA. For the blue branch, a 400
lines/mm grating was used, while the red branch was equipped with a
300 lines/mm grating. The nominal dispersions were 1.83\AA/pixel and
2.31\AA/pixel, respectively, with central wavelengths at 4500\AA~
and~ 7500\AA. The nominal wavelength coverage provided by this
instrument setup around the central value was 4660\AA ~and 7620\AA~
for the two branches, respectively. However, due to vignetting towards
the chip edges, the wavelength regions actually used were somewhat
reduced (blue: 3750 - 5400\AA, red: 5200 - 8000\AA). A 1.5'' slit
was employed.

Exposure times ranged from 900 to 2700~s, depending on the magnitude
of the object, as well as on seeing and weather conditions. At least
three exposures were obtained per object; some of the targets required
4 or 5 exposures. Each on-target series of exposures was followed by a
comparison lamp exposure (HeNeAr) for wavelength calibration.  Spectra
of two spectrophotometric flux standards (Feige110 and BD+28-4211)
were acquired during the same observing runs. A series of 22
spectroscopic standards of mainly early spectral types were also
observed.

\begin{figure}

  \centerline{\psfig{file=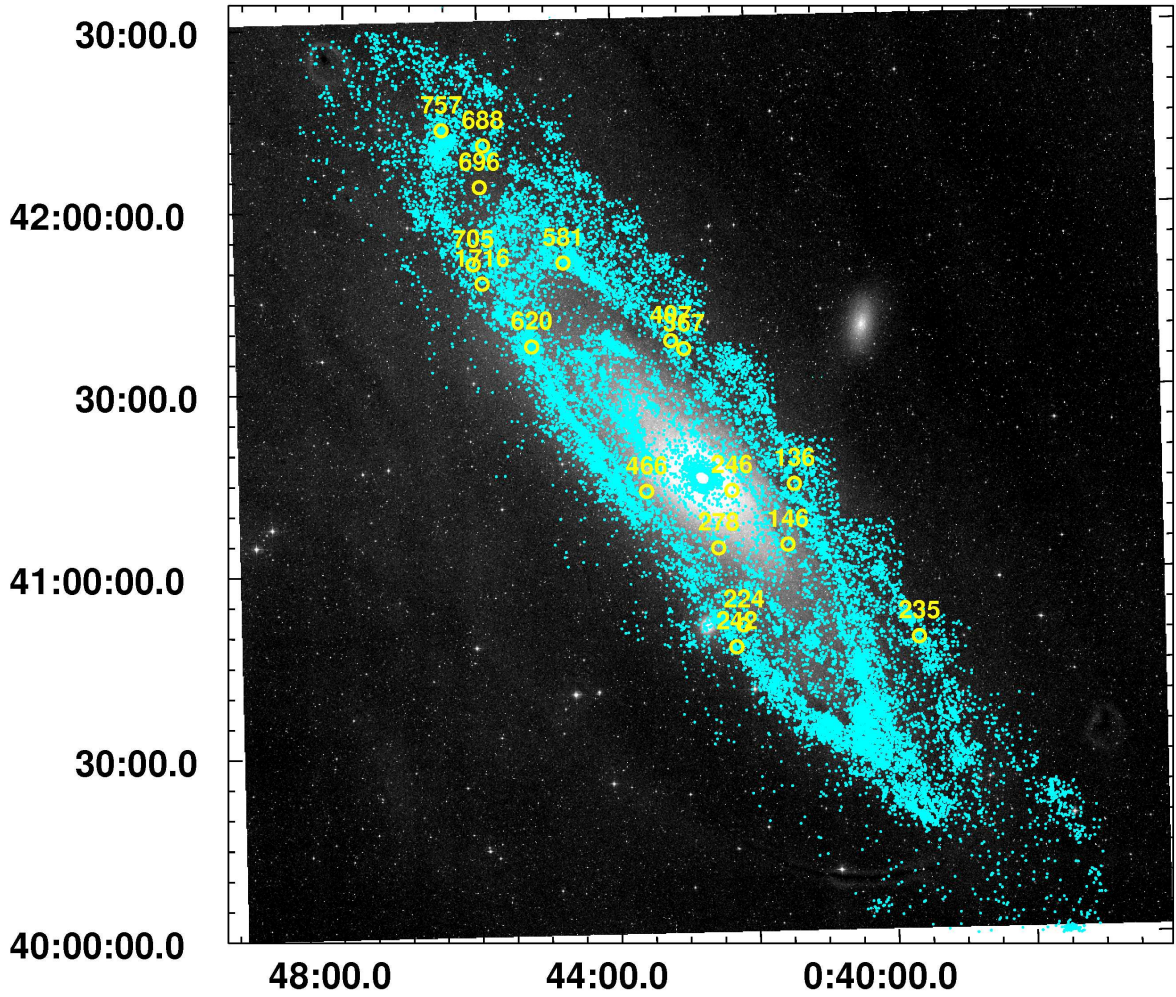,width=6.5in,angle=0}}
  \caption{The location of the observed objects (yellow symbols) on an optical unfiltered image of M31 obtained
  with the Schmidt telescope on Calar Alto (using HDAP which was produced at Landessternwarte Heidelberg-K$\ddot{\rm o}$nigstuhl
  under grant 00.071.2005 of the Klaus-Tschira-Foundation). The locations of OB stars in M31, delineating the spiral
  arms, are marked with cyan.}
  \label{locations}
\end{figure}

Figure~\ref{locations} shows the location of the observed sources (red
symbols) on an optical unfiltered image of M31 obtained with the
Schmidt telescope on Calar Alto (using HDAP which was produced at
Landessternwarte Heidelberg-K\"{o}nigstuhl under grant No.
00.071.2005 of the Klaus-Tschira-Foundation). On the same plot we have
marked with cyan the locations of OB stars in M31 (obtained using
appropriate colour and magnitude cuts from the \citealp{massey2006}
catalogue), which delineate the spiral arms. The vast majority of the
objects were observed both in blue and red, usually with the same
total exposure time.

A total of 20 objects' (non-AGN) spectra have been acquired, as likely
counterparts of 17 X-ray sources. 16 of the spectra obtained have
coverage from 3750-8000~\AA, the others have only 3750-5200~\AA\,
because the red detector was not functioning from November 2006 to
September 2007. The details of the observations of stellar spectra are
reported in Table~\ref{Log}, the X-ray properties of the objects,
including positional offsets, are in Table~\ref{X-ray}, while the LGGS
optical photometry data and 2MASS IR photometry data for the
counterparts observed are given in Table~\ref{Optical}. The other
spectra will be described in a paper detailing the AGN (J. Green et
al., in prep.).  The structure of the Table~\ref{Log} is as follows:

\begin{enumerate}
 \item {\em Column 1} gives the \citet{massey2006} catalogue number
 \item {\em Column 2} gives the X-ray source number of the target as it appears in PFH2005.
 \item {\em Column 3} lists the Right Ascension of the optical counterparts (epoch 2000).
 \item {\em Column 4} gives the Declination of the optical counterparts (epoch 2000)
 \item {\em Column 5} gives the date of the observation at APO telescope.
 \item {\em Column 6} shows whether the target was observed in the blue (B) or red (R) channel
 \item {\em Column 7} lists the total exposure time for the specific spectrum
 \item {\em Column 8} provides the number of individual exposures acquired for the corresponding spectrum.
\end{enumerate}

Finally, we also observed a set of early-type stars with known
spectral types, given in Table~\ref{standards}, for use in
cross-correlation tests to quantify our qualitative classifications
where possible.

\subsection{Data reduction}

The data reduction was performed with the \emph{NOAO-IRAF} FITS Image
Kernel (July 2003), using standard procedures: First, the frames were
bias-subtracted and flat-fielded. The 2-D spectra were subsequently
traced and extracted using the all-in-one package
\emph{apextract}. Comparison lamp spectra were then extracted from the
corresponding lamp exposures, using the profiles measured for the
corresponding object spectra. The comparison lamp spectra were used to
provide the wavelength calibration of the object spectra.  The
resulting accuracy of the wavelength calibration is $\sim 1$\AA (see
\S\ref{observations} for spectral resolution).  Flux calibration was
achieved using the spectrophotometric flux standard spectra.  The
spectrophotometric standard spectra were processed with the same tools
as the object spectra.  Then the fluxes as a function of wavelength
were compared to the published values from \citet{massey2006} to
produce a sensitivity map, which was then applied to all of our
observed spectra. The calibrated object spectra were combined,
weighting according to exposure time, to yield the final spectrum.
This reduction and calibration procedure was also adopted to the
spectroscopic standards.

 \subsection{Spectral Classification}
A first level of classification of the observed spectra was performed
by visual inspection.  The original set of spectra obtained, included
several cases of active galactic nuclei (AGN), showing characteristic,
usually broad, emission lines. These will be discussed in a separate
publication (J. Green et al., in prep.).  The rest of the spectra were
either stellar or nebular (SNRs or HII regions) or a combination of
the two.  Spectra showing a strong continuum with relatively narrow,
or lack of, absorption and/or emission lines, were classified as
stellar. If a spectrum showed strong H$\alpha$ and [S II] emission
lines, it was classified as a SNR or an HII region, depending on the
relative strengths of these lines.  Blue stellar spectra were roughly
classified by eye and put through the IRAF cross-correlation routine
{\tt fxcor} against our set of spectral standards in
Table~\ref{standards}, which returns a peak cross-correlation strength
and best-fit velocity.

\section{Results}

After initial classifications into Star, AGN, or SNR/HII region, we
had 20 stellar spectra, or stellar spectra combined with diffuse
emission, corresponding to 17 X-ray sources, out of spectra of
counterpart candidates of 41 X-ray sources.  Sources [PMH 224],
[Steile 235], and [PMH 367] each have 2 counterpart candidates for
which we obtained spectroscopy.

\begin{figure}
 \centerline{\psfig{file=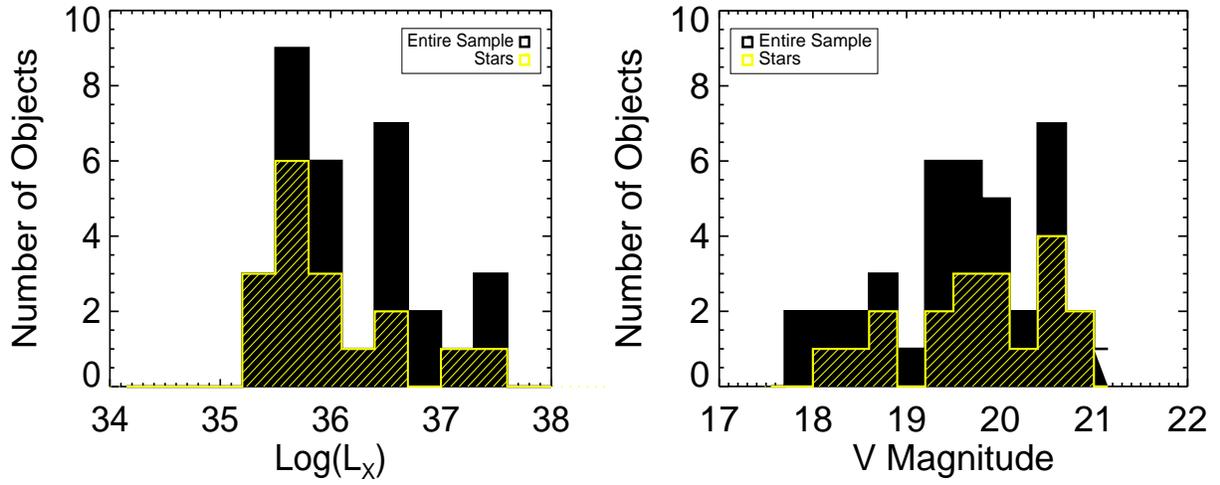,width=6.5in,angle=0}}
  \caption[]{{\it Left:} Distribution of 0.2-4.5 keV X-ray
  luminosities (taking fluxes from \citet{pietsch2005} and
  \citet{stiele2011} and assuming they are at the distance of M31) for
  the sample of sources with optical counterpart candidates observed
  by our study (black), and those that we classified as stars
  (yellow). {\it Right:} Distribution of $V$-band magnitudes of the
  counterpart candidates observed for this study (black), and those
  that we classified as stars (yellow).}
\label{fluxhist}

\end{figure}

The optical and X-ray brightness distribution of the stars, compared
to the overall distribution of our sample is shown in
Figure~\ref{fluxhist}.  We then determined which of these stars were
in M31 using their velocities.  Any star showing an emission or
absorption line feature corresponding with a blue shift corresponding
to the velocity of M31 was considered a M31 member.  Of the 20 spectra
considered here, 2 are foreground objects (LGGS J004123.75+411459.6,
LGGS J004445.06+415153.4), 5 are unlikely to be the correct
counterpart (LGGS~J003937.27+404928.9, LGGS~J004216.78+404814.4,
LGGS~J004259.07+413731.4, LGGS~J004456.78+413548.0,
LGGS~J004604.83+415142.4), two are different parts of the same SNR in
M31 (LGGS~J004210.24+405149.4, LGGS~J004210.38+405148.1). Two are star
clusters in M31 (LGGS J004218.72+411401.2, LGGS J004309.86+411900.9)
The other 9 are stars in M31.

Any star in M31 with luminosities meeting our selection criteria that
is a member of M31 (V$<$20.5; M$_{\rm V}{<}-4$ at the distance of M31)
associated with an X-ray source with a hard spectrum should be
considered a good HMXB candidate; however, relatively red spectra
could be low-mass stars in the late stages of stellar
evolution. Furthermore, stellar spectra showing no atypical features
(e.g., emission lines), may be chance superpositions with the X-ray
source error circle.  Therefore, our best HMXB candidates are those
that are the bluest and show atypical spectral features for single
stars. There are 11 X-ray sources with a counterpart candidate
spectrum of a star in M31, 9 of these 11 have spectral types that only
apply to stars with masses $\gap$5~M$_{\odot}$ (one is a WN star, 8
are HMXB candidates), and 3 of these 9 have spectral features that may
indicate interaction with an X-ray-emitting companion ({\it i.e.,}
emission lines).  We discuss all of the spectra in detail below.

\subsection{[PFH2005] 136 -- Galactic Dwarf Nova}

The only optical counterpart within the positional error-circle of
this X-ray source down to the LGGS \citep{massey2006} limiting
magnitude, is the peculiar variable R066 \citep[][their object
R066==R081]{Rosino1973}. This object is known to show repeated optical
outbursts, and has been classified by Rosino as either a recurrent
nova in M31, or, more likely a U GEM variable \citep[dwarf nova from
accreting white dwarf][]{Tovmasyan1984} in our Galaxy.
\citet{SharovandAlksnis1989} pointed out that a U GEM variable
classification would be consistent with the short interval between
outbursts (recurrent novae flare up more rarely) and the very high
rate of decrease in brightness following an outburst. The object is
also given a U Gem-variable ``uncertain" classification in the Catalogue
of Cataclysmic Variables of \citet[][online edition updated in
2006]{Downes2001}, in the General Catalogue of variable stars
\citep[as M31-V0609;][]{Samus2009}, and in the AAVSO International
Variable Star Index VSX \citep{Watson2013}. 

\begin{figure}
 \centerline{\psfig{file=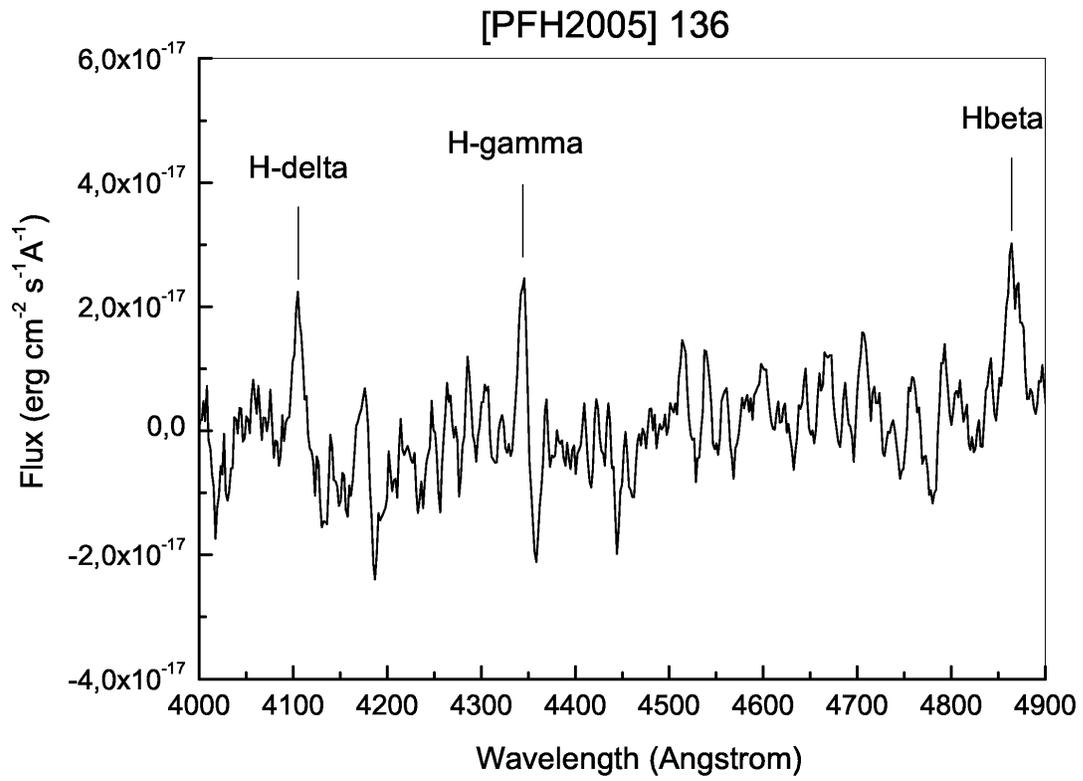,width=6.5in,angle=0}}
  \caption[]{ Blue spectrum of [PFH2005] M31\_136 showing Balmer line emission, expected
  in spectra of UGem variables when in quiescence.}
\label{R066bluespec}
\end{figure}

\begin{figure}
 \centerline{\psfig{file=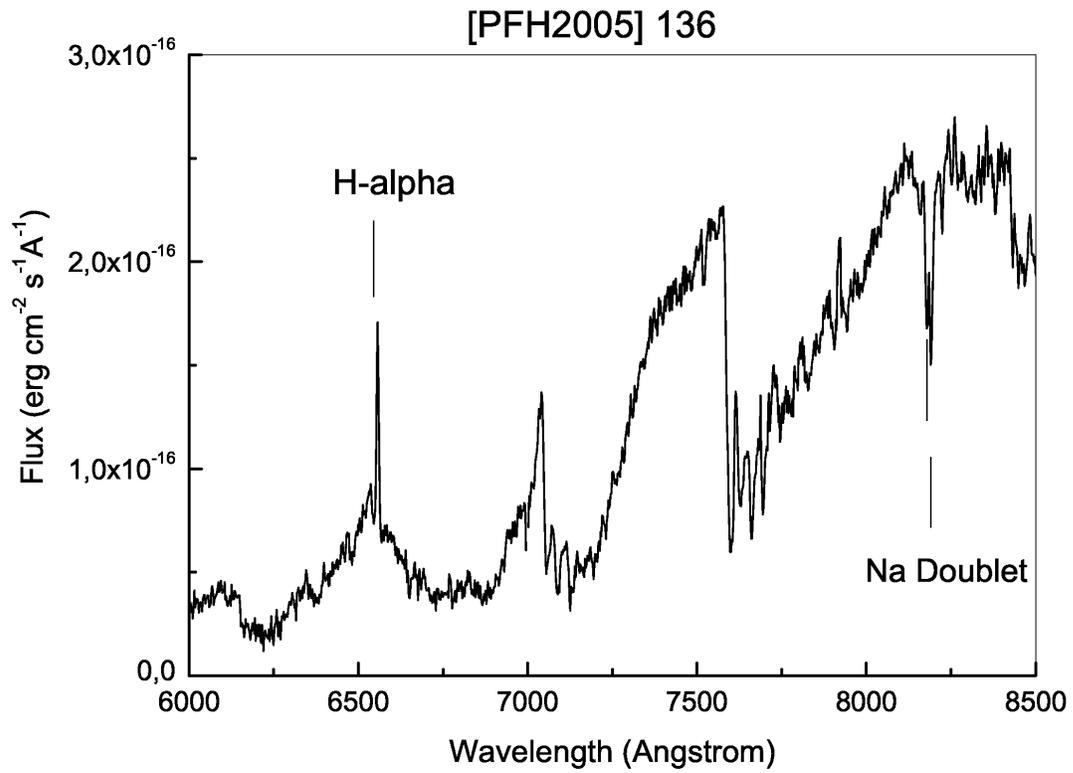,width=6.5in,angle=0}}
  \caption[]{ Red spectrum of [PFH2005] M31\_136 showing the
  characteristic features of an M5-6 dwarf, with H$\alpha$ emission.}
\label{R066redspec}
\end{figure}

The blue stellar spectrum we obtained for R066
(Figure~\ref{R066bluespec}) shows clear Balmer line emission
(H$_\beta$, H$_\gamma$, H$_\delta$ and H$_\epsilon$) with a velocity
of $+$70~km~s$^{-1}$.  The red stellar spectrum
(Figure~\ref{R066redspec}) is consistent with an M5/M6 dwarf in our
Galaxy, with prominent TiO bands (6322\AA, 6569\AA, 6651\AA, 7053\AA,
7666\AA, 8206\AA, 8432\AA). The classification of the star as a dwarf
is based on the strength of the near infrared sodium doublet (Na
8183/8195\AA), which is a sensitive gravity indicator
\citep{Schiavon1997}.  The 2-MASS infrared colours ($J-H=0.47\pm0.09$,
$H-K=0.45\pm0.10$) are also consistent with this
classification. Indeed, at quiescence, U Gem variables (dwarf novae)
are characterized by a strong Balmer emission spectrum on a blue
continuum, while in the infrared, in relatively long orbital period
systems, the spectrum of the secondary dominates. The secondary in
U-Gem's is very often a mid M dwarf, as is the case here
\citep[e.g.][]{Warner1995}.

Most bright dwarf novae at quiescence have been detected in hard
X-rays (0.1-5~keV), while soft X-rays (100-200 eV) have been detected
in quiescence in very few systems \citep[e.g.][]{Warner1995}.  Thus,
both optical and X-ray properties of source [PFH2005] 136 are
consistent with it being a dwarf nova in the Galaxy.

\subsection{[PFH2005] 146 -- M31 WR and O Star Bubble?}

Within the 2.5-sigma error circle (3.65 arcsec radius) of the X-ray
position, there are 5 candidate optical counterparts in the Massey
catalogue.  We have obtained a spectrum of the brightest and bluest
counterpart, LGGS\_J004130.37+410500.9.  The others are separated by
$>$1$''$, and are fainter, so that they are unlikely to have
significantly affected the spectrum of LGGS\_J004130.37+410500.9.

The X-ray source [PFH2005] 146 is spatially correlated with a
catalogued HII region in M31
\citep{BaadeandArp1964,Pellet1978,Kraemer2002}.  The object is also
recorded in the radio in \citet{Braun1990}.  PFH2005 have classified
it as an SNR, on the basis of its hardness ratios and the presence of
radio emission.  The LGGS object we observed is catalogued as an
H$\alpha$ emission line object in \citealp{massey2006} who have
classified it as a Wolf-Rayet star of WNL type embedded in an HII
region, on the basis of a low resolution spectrum.

\begin{figure}
 \centerline{\psfig{file=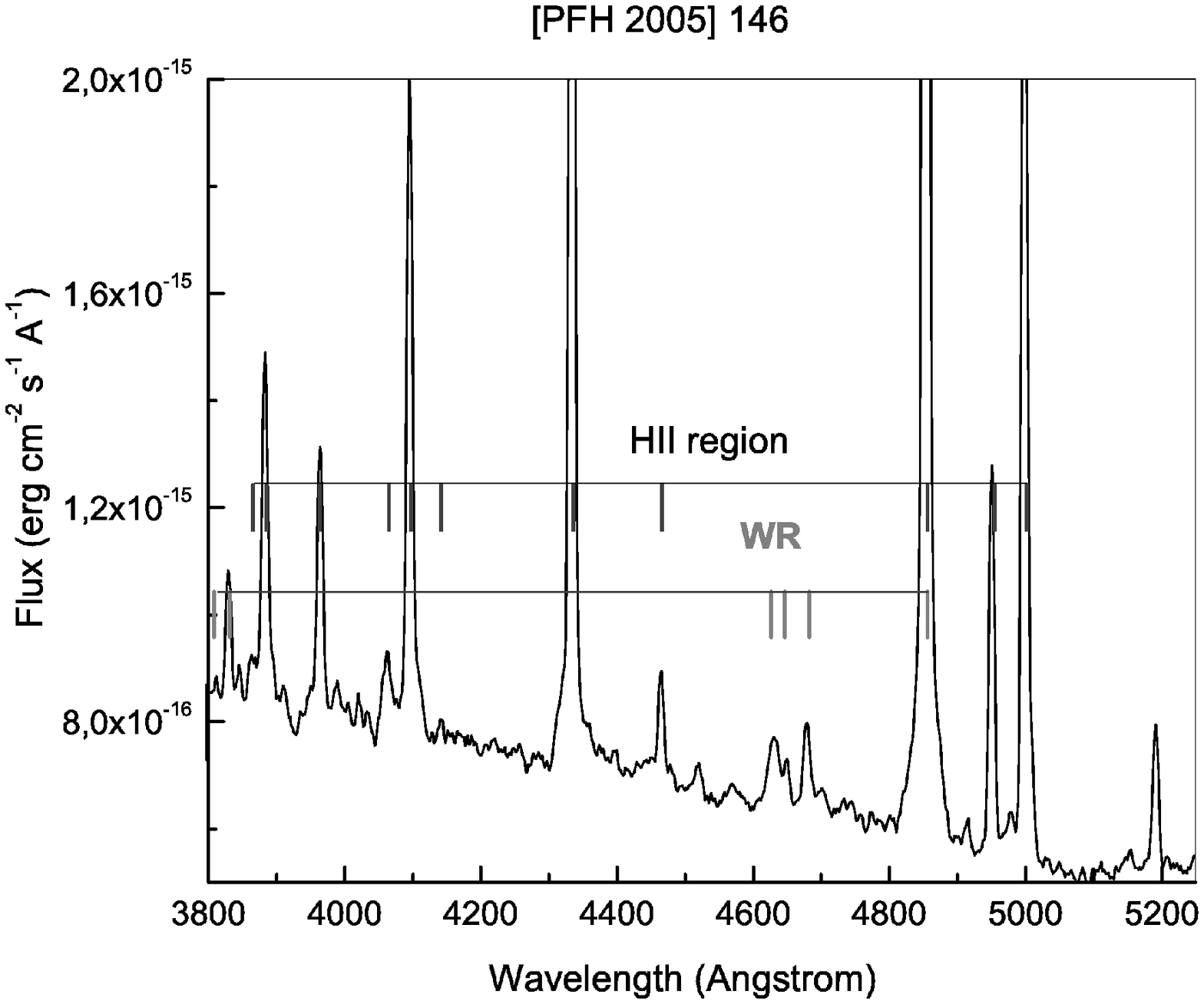,width=4.5in,angle=0}}
  \caption{ Blue part of the spectrum of [PFH2005] M31\_146 showing characteristic lines of a WN Wolf Rayet star embedded in an HII region. }
\label{146_blue}
\end{figure}

\begin{figure}
 \centerline{\psfig{file=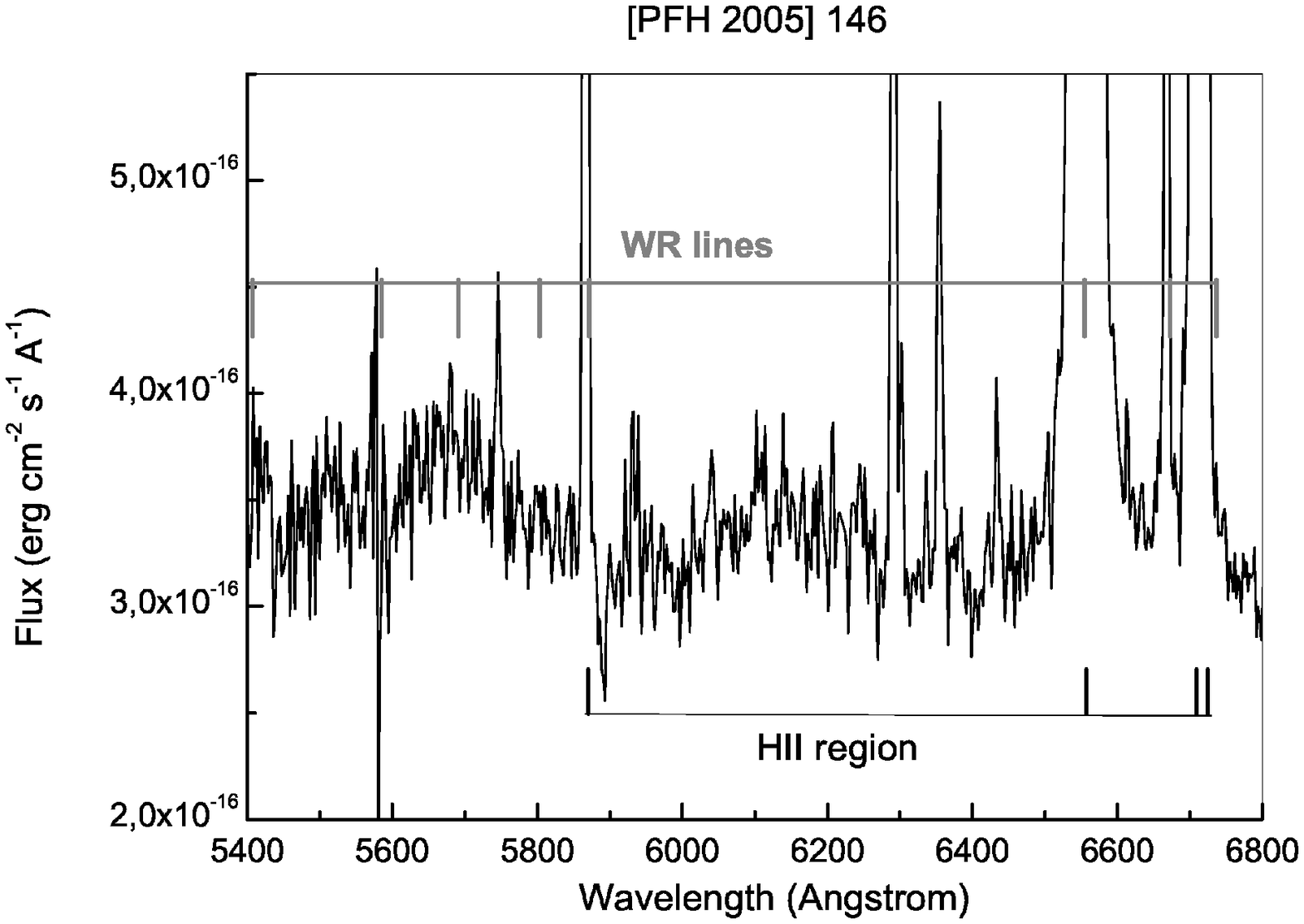,width=4.5in,angle=0}}
  \caption{ Red part of the spectrum of [PFH2005] M31\_146 showing characteristic lines of a WN Wolf Rayet star embedded in an HII region. }
\label{146_red}
\end{figure}

Our spectrum of LGGS\_J004130.37+410500.9 is very similar to the
spectrum of NGC604-WR12 in M33 of type WN, in \citet{Drissen2008}.  It
appears to be a composite spectrum of a WR star and an HII region. In
Figures~\ref{146_blue} and \ref{146_red}, we label parts of the
spectrum with the characteristic lines of HII regions and WR stars
identified in different colours.

\subsubsection{HII- region component}

Table~\ref{146_lines} presents a summary of the most significant
 nebular emission lines identified in the spectrum of the object
 [PFH2005] 146. We report the rest frame wavelengths and the observed
 ones, as well as the fluxes derived using the deblending/fitting
 function provided by the \emph{IRAF} task noao.onedspec.\emph{splot}.

The blueshift of the lines is $\Delta\lambda\simeq$10\AA,~which yields an
approximate radial velocity of $RV\sim-450$~km~s$^{-1}$, thus
confirming M31 membership.

The line ratios of characteristic lines are compatible with an HII
region. SNR shocks are expected to create strong $[SII]$
$\lambda\lambda$6716,6731 emission lines, typically with
$[SII]/H{\alpha} > 0.4-0.5$ - empirical criterion value (see for
example \citealp{MatonickandFesen1997}, and references
therein). Referring to Table \ref{146_lines}, for object [PFH2005] 146
this ratio is about 0.29.  Furthermore, the relative fluxes
$[SII](6731+6716)/H{\alpha}$ and $[OIII](5007)/H{\beta}$ for the the
object data are well within the HII region boundary of
\citet{Kewley2001}.

\subsubsection{Wolf-Rayet stellar component}

The emission lines from the Wolf-Rayet component of the spectrum are
summarized in Table~\ref{146_lines_wr}. We used the complete line list
for HII regions from \citet{garcia-rojas2006}, to separate out the HII lines.
 The optical spectrum
we obtained shows strong emission in HeII 4686 and in NIII and
NIV. These are characteristic of WN stars \citep[see
e.g.][]{Niemela2006}.

The X-ray emission with f$_{\rm X}=2.6{\times}10^{-15}$ mW~m$^{-2}$
has a soft spectrum (HR1=$-$0.97), which would be consistent with a
SNR. However, the optical spectrum clearly indicates an HII region.
There are wind-blown bubbles around O, B or WR stars, which are
related to HII regions in that they are also regions of ionized gas
surrounding a massive star \citep[see][]{Magnier1995}. Wind-blown
bubbles, from those around massive O and Wolf-Rayet stars to
superbubbles around OB associations and Galactic winds in starburst
galaxies, are X-ray emitters as most of their volume is taken up with
hot gas, $10^5 \leq T(K)\leq 10^8$. \citet{StricklandandStevens1998}
predict X-ray emission from wind-blown bubbles, which is soft but with
possibly a hard tail.

Wolf-Rayet stars that are bright in X-rays are nearly always members
of binary systems \citep{GuerreroandChu2005}.  Many Wolf Rayet stars
are binary systems. These often occur as W + O systems. As the O star
plows through the thick wind from the WR star, a shock wave can be
formed, often emitting X-rays.  The X-ray survey of WR stars in the
MCs has shown that WR stars in binary systems are much more frequently
detected in X rays than single WR stars and that they have higher
L$_{\rm X}$/L$_{bol}$ than apparently single WR stars. The X-ray
luminosities of these X-ray-bright WR binaries are most likely
enhanced by the colliding stellar winds: In a WR+OB binary system, the
WR wind collides with the companion's fast wind and generates
shock-heated plasma at the interaction region. The X-ray luminosities
of the WR stars in binaries the MCs span a wide range, from a few
$\times 10^{32}$ erg~s$^{-1}$ to $\simeq2\times 10^{35}$ erg~s$^{-1}$
\citep{GuerreroandChu2005}.

Source M31\_146 is just above the detection limit of the X-ray survey
(10$^{35}$ erg s$^{-1}$). It also has a soft spectrum.  The combined
X-ray and optical properties of the source are compatible with a WN
and an O star binary, surrounded by an ionized bubble.

\subsection{[PFH2005] 224 -- M31 Supernova Remnant}

\citet{pietsch2005} classify this object as a supersoft source
(although \citealp{stiele2011} do not).  There are five candidate
optical counterparts in LGGS (\citet{massey2006}), four of which are
very faint.  On the H$\alpha$ image there is clear indication of an
SNR (ringlike H$\alpha$ structure of 6 arcsec radius, with bright
central spot). However, there is no previously published SNR
\citep{Blair1981,Magnier1995,sasaki2012} at this location. On the
other hand, \citet{Azimlu2011} record 3 HII regions within the error
circle (IDs 1546, 1549, 1547).

\begin{figure}
 \centerline{\psfig{file=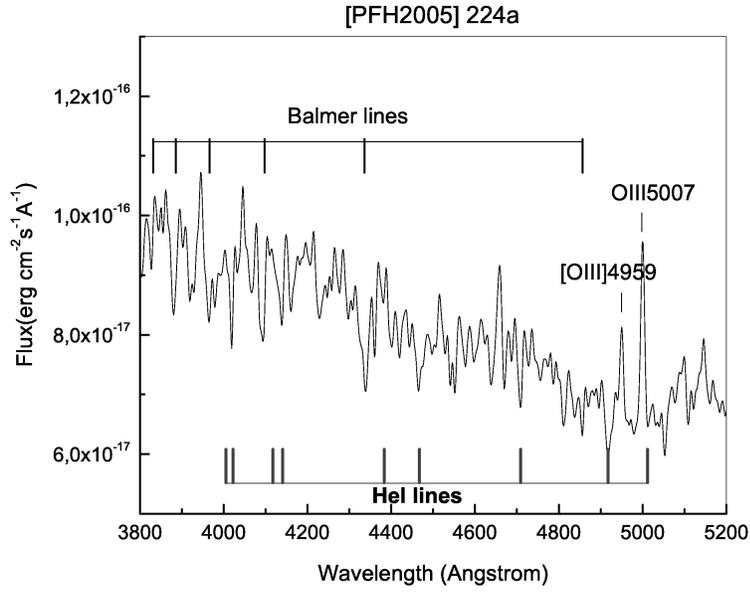,width=4.5in,angle=0}}
  \caption[]{ The blue part of the spectrum of [PFH2005] M31\_224a showing characteristic
  absorption lines of of HeI and Hydrogen characteristic of an OB star. }
\label{224aspec1}
\end{figure}

\begin{figure}
 \centerline{\psfig{file=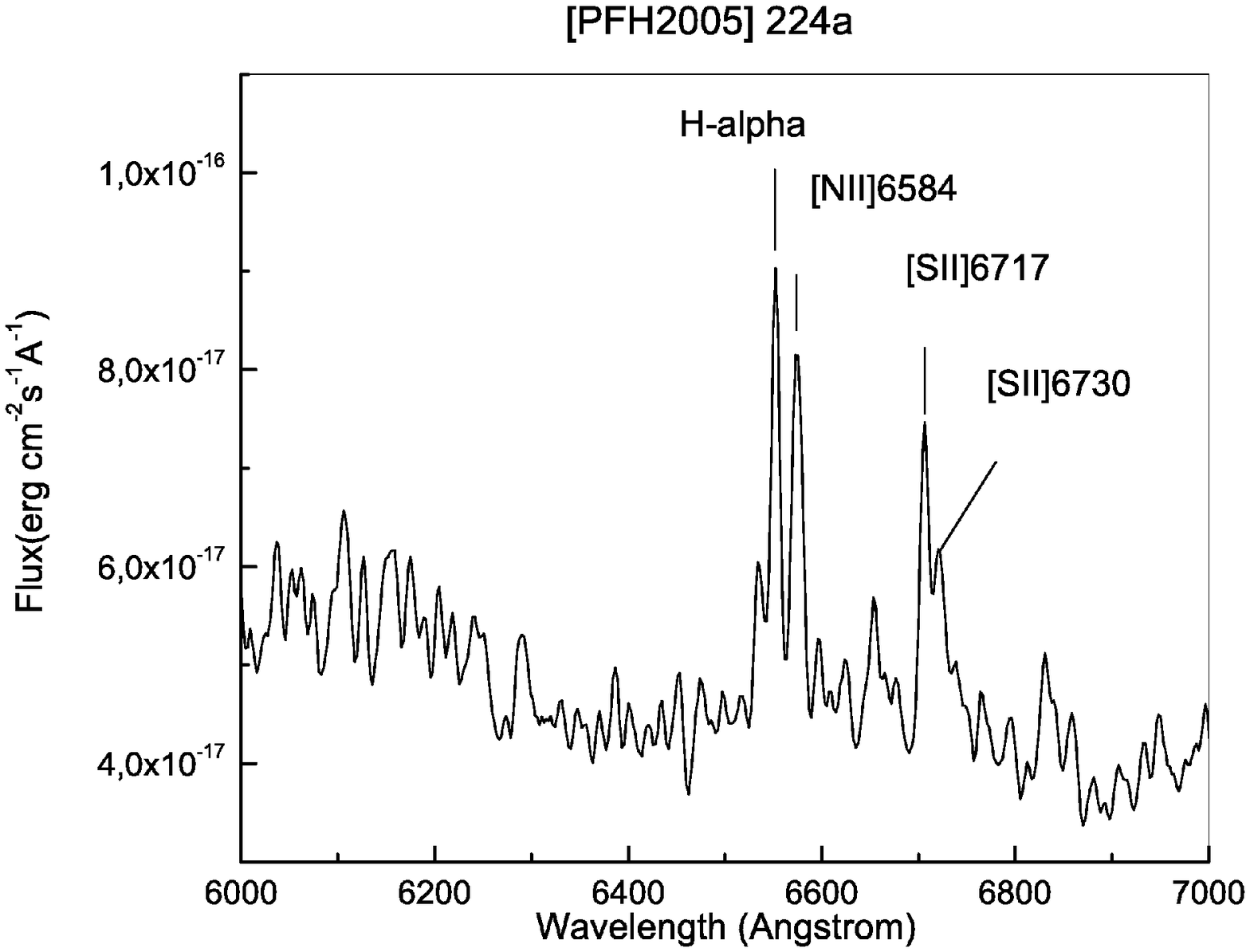,width=4.5in,angle=0}}
  \caption[]{ The red part of the spectrum of [PFH2005] M31\_224a showing characteristic
  emission lines of a SNR. }
\label{224aspec2}
\end{figure}

We obtained spectra of both the brightest central candidate
counterpart (LGGS\_J004210.24+405149.4) and of the faint adjacent
object (LGGS\_J004210.38+405148.1).  The optical spectrum of
LGGS\_J004210.24+405149.4, referred to as object 224a,
(Figures~\ref{224aspec1} and \ref{224aspec2}) has the characteristic
emission lines with the expected line ratios (see description of
object [PFH2005]146) for a SNR at the characteristic blueshift of M31,
$[SII]/H\alpha=1.1$ and $[OIII]/H\alpha=0.7$.  We note that there are
also weak Balmer lines and strong HeI lines in absorption,
corresponding to an early type OB star.  Therefore, there is light
from an OB star contaminating the SNR spectrum. However, the line
ratios and the soft X-ray characteristics confirms that the
X-ray-emitting object is an SNR.  It must be noted that a revised
reduction of the XMM-Newton data \citep{stiele2011} yielded hardness
ratios pointing to a harder source, although the errors are large. In
an attempt to clarify the issue, we have searched in the $Chandra$
archive for possible further information on this source. The source is
visible on an ACIS-S (CCD no.6) 54ks exposure, as a faint soft source,
with most of the photons below 1keV. The specific observation is far
off axis, so it is not possible to determine if the source is a point
source or extended. Therefore the $Chandra$ observations seem to
confirm that the source is soft rather than hard.

\begin{figure}
 \centerline{\psfig{file=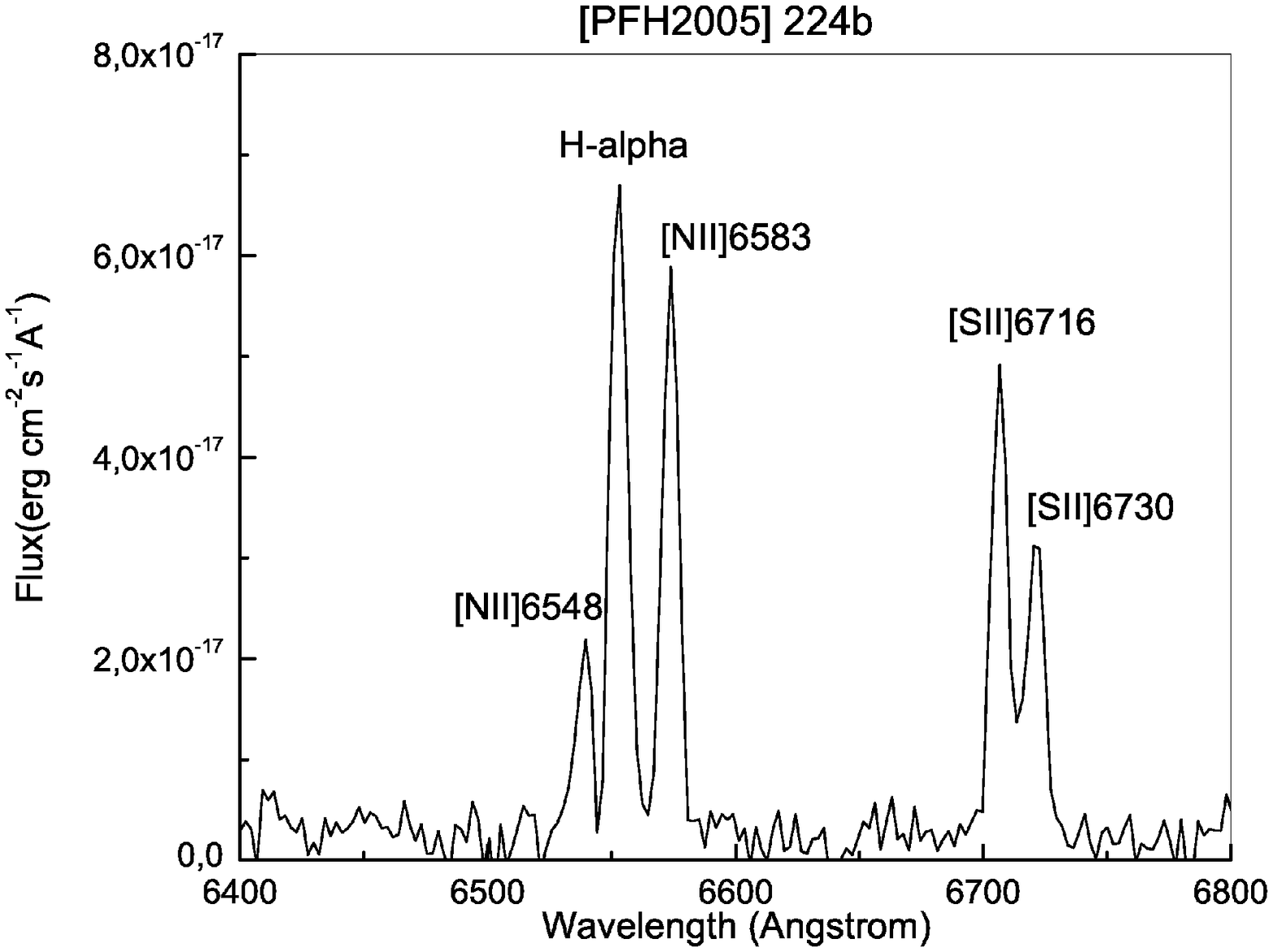,width=4.5in,angle=0}}
  \caption[]{ The red part of the spectrum of [PFH2005] M31\_224b showing characteristic
  emission lines of a SNR. }
\label{224bspec}
\end{figure}

The spectrum of the second object, referred to as object 224b
(LGGS\_J004210.38+405148.1, Figure \ref{224bspec}) shows only SNR
emission lines at the blueshift of M31, with no visible stellar
continuum.

In conclusion, the X-ray emission seems to be related to a newly
identified SNR or superbubble, which is also positionally coincident
with an OB star in M31. Single OB stars have been detected in X-rays
and almost always show soft emission at characteristic temperatures
kT$<$1 keV and low X-ray luminosities \citep{berghofer1997}, well
below the detections limits of the{\it XMM-Newton} survey.

\subsection{[Stiele2011] 235 -- Strong M31 HMXB Candidate}

There are two LGGS objects within the error circle of the X-ray
position, LGGS\_J003937.29+404925.8 and LGGS\_J003937.27+404928.9,
both blue in colour. We have obtained spectra of both objects.

\subsubsection{LGGS\_J003937.29+404925.8 -- Emission line star}

\begin{figure}
 \centerline{\psfig{file=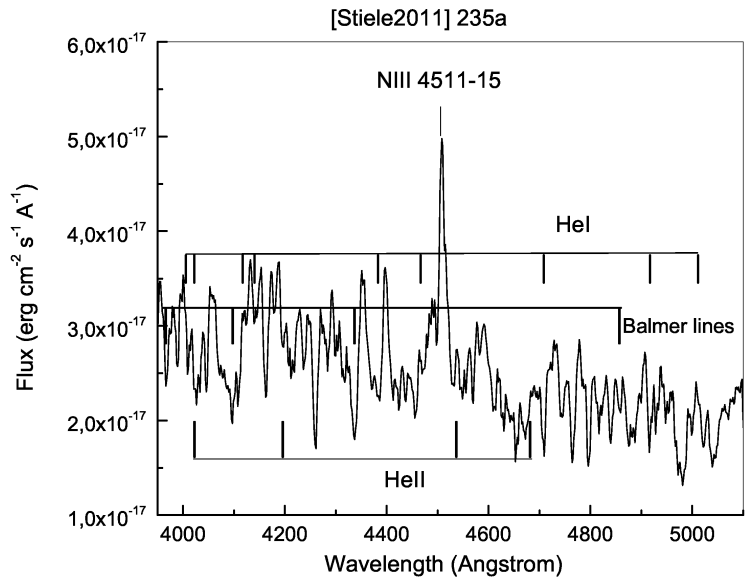,width=4.5in,angle=0}}
  \caption[]{ The blue part of the spectrum of [Stiele2011] M31\_235a
  showing characteristic emission of an O-type star. }
\label{235aspec1}
\end{figure}

\begin{figure}
 \centerline{\psfig{file=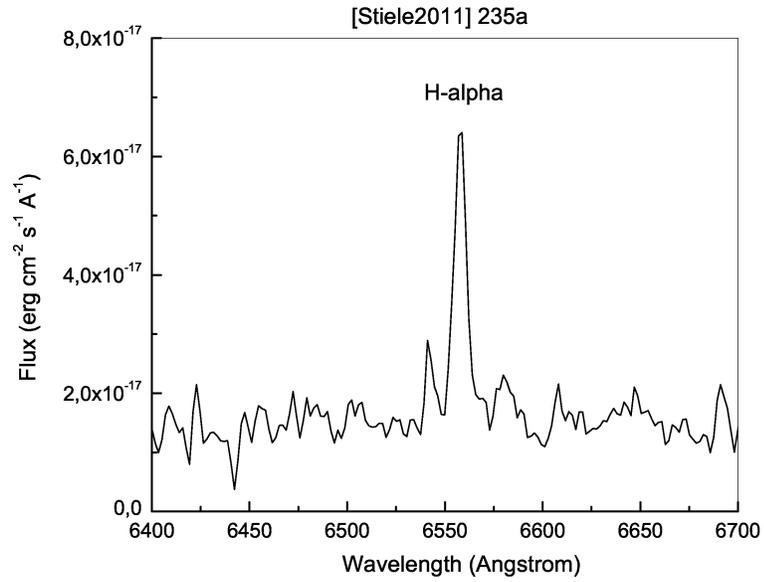,width=4.5in,angle=0}}
  \caption[]{ The red portion of the spectrum of [Stiele2011] M31\_235a
  showing a strong H$\alpha$ emission line.}
\label{235aspec2}
\end{figure}

The blue and red spectra of LGGS\_J003937.29+404925.8, the brightest
 and bluest of the two candidate counterparts, are shown in
 Figures~\ref{235aspec1} and \ref{235aspec2}, respectively. The
 spectra are blue-shifted by about -250 km~s$^{-1}$, indicating M31
 membership.  The blue part of the spectrum shows a relatively
 featureless continuum with very weak Balmer absorption and HeI
 absorption at 4387\AA, almost as deep as H$\gamma$.
 Cross-correlation against our spectral standards found no correlation
 with our templates, likely due to the low S/N of the features.  We
 have marked on the spectrum the location of the Balmer, HeI and HeII
 lines expected in early type (OB) stars. However, the S/N ratio is
 insufficient to identify with certainty any other absorption
 lines. There is also relatively strong emission at 4508\AA, which is
 consistent with NIII (4511-4515\AA), often seen in O stars. The
 weakness of the Balmer lines, the presence of some HeI in absorption
 and NIII in emission suggest a late O star.  Using LGGS photometric
 data we derive a value of $\simeq$-0.92 for the reddening free
 Q-parameter, which is also consistent with an O9-B0 star, in
 agreement with our qualitative spectral classification. The derived
 Q-value is also well within the known range of Q values for Be stars
 (in the Galaxy the Q values of Be stars range between $-0.322$ and
 $-1.024$).

The red part of the spectrum shows clear H$\alpha$ emission, with a
FWHM of $\simeq 7.7$\AA, or $\simeq 5.7$\AA\ (corresponding to
$\simeq240~km~s^{-1}$), after correcting for the spectral resolution.
This value is quite normal for BeXRBs (e.g. BeXRBs in SMC cluster
NGC330 \citealp{martayan2007}).  6560\AA) Taking all the evidence into
account, the star is of a late O-early B spectral type with apparently
associated H$\alpha$ emission. 

In X-rays, the source is classified as a hard source, and its flux
corresponds to an X-ray luminosity of
$L_{\rm X}=4.7\times10^{37}$ erg~s$^{-1}$. Most BeXRBs have X-ray luminosities
$\leq3\times10^{36}$ erg~s$^{-1}$, but a few have been recorded up to
$10^{38}$ erg~s$^{-1}$.

The combination of optical and X-ray data make this object
(LGGS\_J003937.29+404925.8) a very good HMXB candidate in M31,
possibly a BeXRB.

There is however some evidence that reduces somewhat the confidence
level of this classification. The H$\alpha$ emission observed may be
partly due to diffuse emission (from an HII region, not an SNR,
because there is no [SII] visible in the spectrum): There is some
indication of very weak diffuse H$\alpha$ emission in the LGGS images
around the position of LGGS\_J003937.29+404925.8.  Moreover the
spectrum shows some weak [NII] 6548\AA\ emission which is consistent
with nebular emission. Additionally, the second candidate counterpart
discussed below also shows similar emission in H$\alpha$. Therefore, at
least part of the H$\alpha$ emission recorded maybe due to a small
surrounding nebula.

\subsubsection{LGGS\_J003937.27+404928.9 -- Early-type Star}

\begin{figure}
 \centerline{\psfig{file=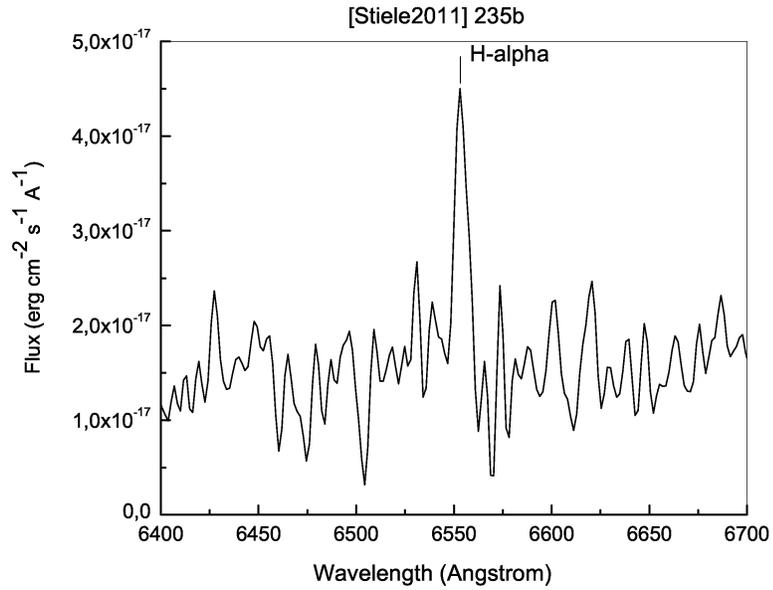,width=4.5in,angle=0}}
  \caption[]{ The red portion of the spectrum of [Stiele2011] M31\_235b
  showing H$\alpha$ in emission.}
\label{235bspec}
\end{figure}

We have also obtained spectra for the fainter candidate optical
counterpart LGGS\_J003937.27+404928.9.  The blue spectrum is dominated
by noise, with no spectral features identifiable. The red spectrum
(\ref{235bspec}) shows H$\alpha$ emission, with similar FWHM as in the
case of LGGS\_J003937.29+404925.8.  Using LGGS photometric data we
derive a value of Q$\simeq$-0.83 for the reddening free Q-parameter,
which is consistent with a B0 star.  Although the S/N of the spectra
did not allow the spectral classification of
LGGS\_J003937.27+404928.9, we cannot rule out the possibility that
this star is the counterpart of the X-ray source. In this case as
well, the object would be a good HMXB candidate.

\subsection{[PFH2005] 242 -- Foreground Star}

This source has a soft X-ray spectrum in \citet{stiele2011}, making it
unlikely to be an HMXB, but its bright blue counterpart candidate put
it into our sample. There are four objects recorded in the Massey
catalogue within the error circle of the revised X-ray position
\citep{stiele2011}, which will be discussed separately below.

\subsubsection{LGGS\_J004216.78+404814.4 -- Chance Superposition}

\begin{figure}
 \centerline{\psfig{file=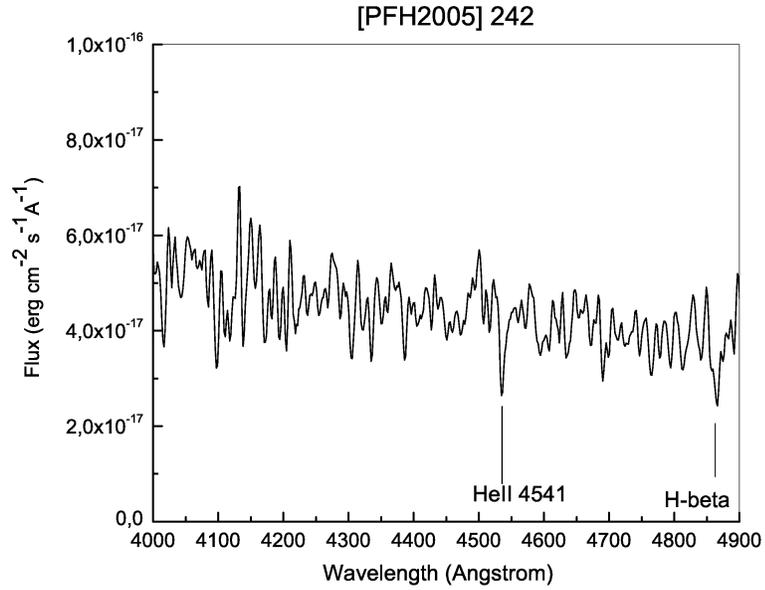,width=4.5in,angle=0}}
  \caption[]{Blue spectrum of [PFH2005] M31\_242 showing possible HeII line}
\label{242spec}
\end{figure}

This is one of the two brighter objects in the error circle, for which
we have obtained an optical spectrum.  The magnitude and colours
($V=19.419\pm0.004$ $B-V=-0.104\pm0.005$, $U-B=-0.993\pm0.004$, $V-R=
-0.013\pm0.006$ $R-I=-0.010\pm0.005$) are consistent with an OB dwarf
in M31. The spectrum obtained (Figure~\ref{242spec}) is almost
entirely featureless, indicative of an OB dwarf at this
luminosity. The H$\alpha$ line appears in absorption at 6554\AA,
blue-shifted by about 410 km~s$^{-1}$. There is a weak HeI 6678 line
appearing at 6669\AA\ at the same blueshift as H$\alpha$, which
confirms its identification.  There is also a marginal detection of
HeII ($\lambda$4541) line at the same blueshift (observed at
$\simeq$4535\AA). These line identifications confirm that the star is
of 09-BO spectral type. The spectral type is also consistent with the
value of the reddening free Q parameter
(Q$_{UBV}=-0.918\pm0.006$). The star is also on the locus of Be stars
on the U-B vs B-V diagram \citep{FeinsteinandMarraco1979}. However,
the hardness ratios obtained for the X-ray source are inconsistent
with the BeXRB.  Actually, they are more consistent with a foreground
star ({\it cf.} Table~\ref{X-ray}).

In conclusion, the observed brightest optical counterpart of [PFH
2005] 242 is an early type OB star in M31, with no H$\alpha$ emission.
The inconsistency of the X-ray hardness ratios with a HMXB render
the probability that this is the correct counterpart of the X-ray
source low. Therefore we consider the rest of the optical sources in
the X-ray error circle, in the following subsections.

\subsubsection{LGGS~J004216.61+404815.9 -- Chance Superposition}

This object is much fainter, very red and probably extended.
$V=22.602\pm0.061$ $B-V=1.817\pm0.477$, $V-R=0.895\pm0.071$,
$R-I=0.911\pm0.037$.  Assuming that the object is in M31
($m-M_0$=24.47, and $E(B-V)=0.08$ \citealp{mcconnachie2005}), we would
have $M_V\simeq-1.9$, $(B-V)_o\simeq1.74$, $(V-I)_o\simeq1.7$, which
are not consistent with any stellar object in M31.

\subsubsection{LGGS\_J004216.47+404812.0 -- Foreground Star}
This object is red ($V=19.843 \pm0.004$, $B-V=1.575\pm0.011$
$U-B=1.126\pm0.024$ $V-R=1.007\pm 0.005$ $R-I=1.114\pm0.004$), with
colours consistent with late K -early M star. This may be a Galactic
foreground object with proper motion pm$_{\rm RA}$$=-41.5\pm5.7$
mas~yr$^{-1}$ pm$_{\rm DEC}$$=-41.3\pm5.7$mas~yr$^{-1}$ \citep{Roeser2010}
and IR photometry from 2MASS 00421645+4048113 J=16.381(0.117). This
foreground star could be the counterpart of the X-ray source.

\subsubsection{LGGS\_J004216.52+404810.2 -- Foreground Star}
 This object is also red ($V=22.013\pm0.020$,
$B-V=1.531\pm0.125$, $U-B=99.999\pm99.999$, $V-R=0.875\pm0.024$,
$R-I=1.030\pm0.013$). The colours are consistent with a Galactic
M-dwarf, also making this star a possible counterpart candidate.

Therefore, most probably this X-ray source can be identified  with a
foreground M star.

\subsection{[PFH2005] 246 -- M31 Globular Cluster LMXB}

\begin{figure}
\centerline{\psfig{file=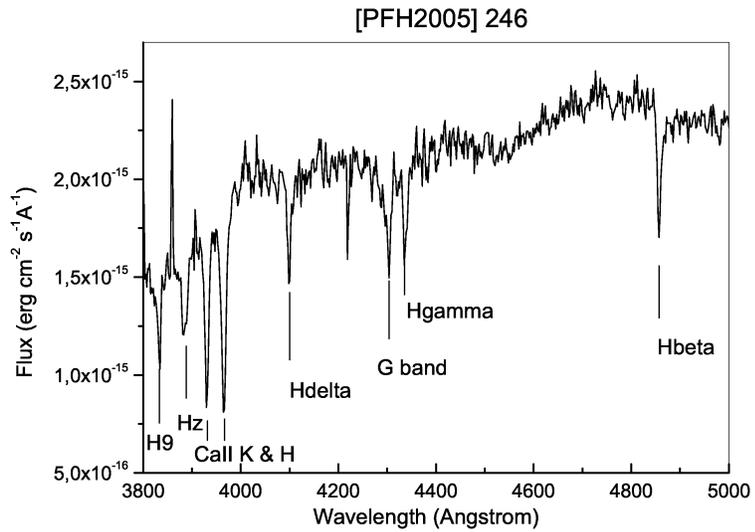,width=4.5in,angle=0}}
  \caption[]{ Blue part of the spectrum of [PFH2005] M31\_246. Composite spectrum of an old globular cluster. }
\label{246spec}
\end{figure}

According to a recent catalogue of globular cluster GCs in M31
\citep{Caldwell2009}, at the position of the X-ray source, there is an
old GC named B086-G148 (0:42:18.65, 41:14:02.1).  The X-ray source has
been widely discussed in the literature
\citep[e.g.]{trinchieri1991,primini1993,supper1997,voss2007}. The
optical spectrum (Figure~\ref{246spec}) is consistent with that of a
GC, with characteristic Balmer lines, H and K CaII and G-band, as
expected from the composite spectrum of a GC
\citep[e.g.,][]{Galleti2007}.  Therefore, this is most likely an LMXB
in the GC.

\subsection{[PFH2005] 278 -- Strong M31 HMXB Candidate}

\begin{figure}
\centerline{\psfig{file=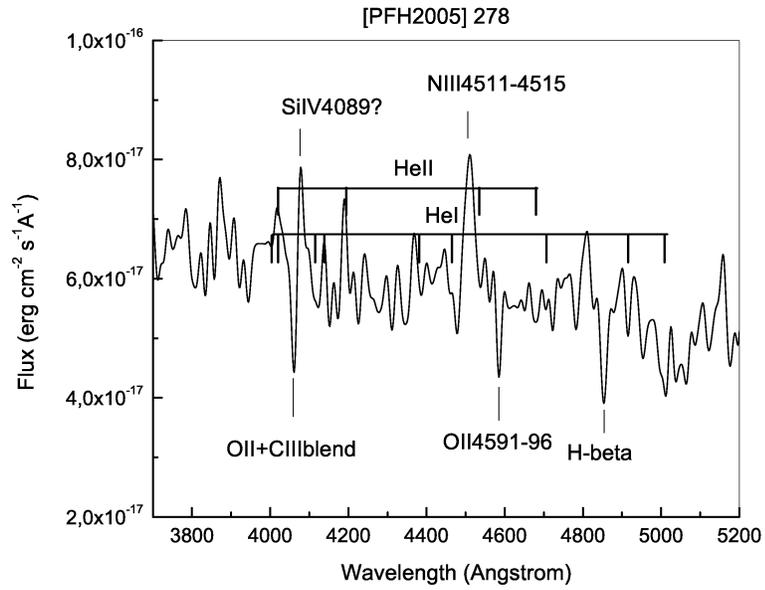,width=4.5in,angle=0}}
   \caption[]{ Part of the blue spectrum of the early type star (early B supergiant)
   which is the brightest candidate counterpart for M31\_278.}
\label{278spec}
\end{figure}

The optical counterpart candidate observed is
LGGS\_J004231.23+410435.3.  It is the brightest star in the LGGS
catalogue lying within the 3-sigma error circle of the X-ray
position. There is one other faint object (LGGS\_J004231.38+410436.1)
within the error circle which will be discussed as well.  The optical
spectrum of LGGS\_J004231.23+410435.3 (Figure~\ref{278spec}) has
relatively low S/N, resulting in no conclusive cross-correlation
results, but some lines can be identified at low significance; the
strongest feature is H${\beta}$ at the blueshift of M31 at ${\simeq}-$400
km s$^{-1}$. At about the same blueshift there are some HeI and weak
HeII absorption lines, as well as OII in absorption, and probable
emission in SiV and NIII. Despite the low S/N the consistency of the
blueshifts of the tentative lines increase confidence in their
identification.  The red part of the spectrum is of low S/N because
the star is blue, and detects no features.  Nonetheless, because the
H$\alpha$ line is not seen either in emission or in absorption, it may
be filled-in by emission.

The identified lines are characteristic of an early type star around
B0, while the observed emission lines are most probably indicative of
a supergiant, although they are not typical. The magnitude and colours
of the star given in LGGS are consistent with a B0-B0.5 II-Ib
supergiant, in agreement with the spectral classification. Thus this
is a high-mass star in M31 with some interesting emission.
Furthermore, it is reported as a variable source in the POINT-AGAPE
survey \citep[Period=10$^3.080$ days (with a false alarm probability
of 10$^-32.9$),][]{An2004}. The X-ray flux of the source at the
distance of M31 is $\simeq1.7\times10^{36}$ erg s$^{-1}$, while it is
classified as a hard source in both PFH2005 and
\citet{stiele2011}. All of these characteristics make [PFH2005] 278 a
good HMXB candidate, probably a SG-XRB.

It must be noted that there is another much fainter object ($V=22.07$)
recorded in LGGS (LGGS\_J004231.38+410436.1) within the error
circle. The object is rather red ($B-V=1.43\pm0.21$), consistent with
a K-M type star, probably a foreground dwarf. The X-ray properties of
the X-ray source are not consistent with a $<$fg$>$ classification;
therefore, this fainter object is not a probable counterpart for the
X-ray source.

\subsection{[PFH2005] 367 -- M31 HMXB Candidate}

Within the [PFH2005] 3-sigma error circle of the source there are
three sources recorded in the \citet{massey2006} catalogue, namely
objects LGGS\_J004258.94+413727.3, LGGS\_J004259.07+413731.4 and
LGGS\_J004259.02+413730.9.  We obtained spectra for the first and
second object, as the third one is too faint ($V=21.55$).

\subsubsection{LGGS\_J004258.94+413727.3 -- OB Supergiant}

This is the brightest and bluest of the objects within the 3-sigma
error circle of the X-ray source.

\begin{figure}
\centerline{\psfig{file=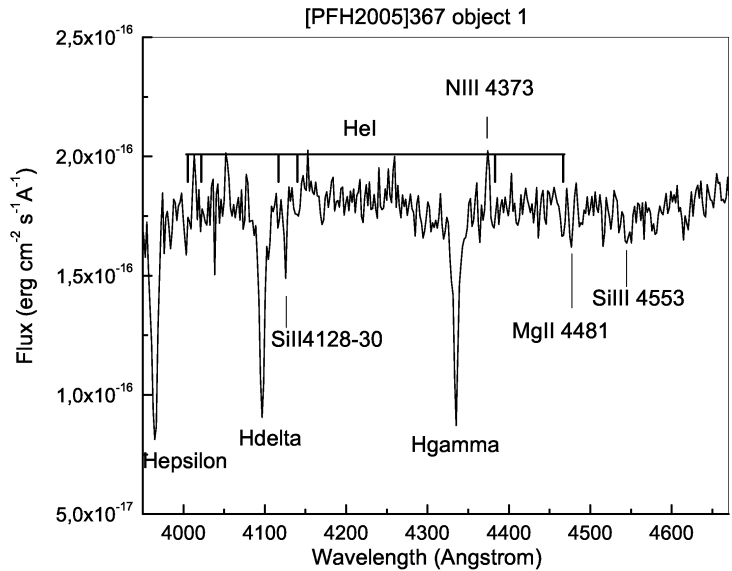,width=4.5in,angle=0}}
   \caption[]{ Part of the blue spectrum of the early type star
   which is the brightest candidate counterpart for M31\_367.}
\label{367aspec}
\end{figure}

The spectrum obtained is contaminated somewhat by
LGGS\_J004258.99+413726.8 (just outside the error circle, 0.75$''$
from LGGS\_J004258.94+413727.3), which is however 1 mag fainter (but
also blue), therefore the prevailing spectrum is that of
LGGS\_J004258.94+413727.3.  The spectrum is shown in
Figure~\ref{367aspec}.  The Balmer lines appear in absorption, at a
velocity of ${\simeq}{-}$320~km~s$^{-1}$, compatible with M31
membership. The spectrum cross-correlates with VOR+59 97 (B2 V) at a
velocity of $-$390$\pm$110~km~s$^{-1}$, and with HD 13799 (B6 III) at
a velocity of $-$350$\pm$60~km~s$^{-1}$.  There are clear He I lines
as well as MgII (4481\AA) and SiII (4128\AA). Using the criteria of
\citet{evans2004} the star appears to be a late B star, probably
B9. However, the strength of the SiIII (4553\AA) line in comparison to
the MgII4481 line is pointing to an earlier type, around B3.  There is
also weak emission of NIII (4373\AA) which is seen in O stars.
However, there are no HeII lines anywhere in the spectrum, as would be
expected for an O star.  In the red part of the spectrum the only
clearly identifiable line is H${\alpha}$ in absorption.  The reddening
free Q-parameter of the star on the basis of the LGGS colours is
Q$=-1.19$, which is not compatible with a late B star, but rather with
an O star. The magnitude of the star at the distance of M31 (and
accounting for the average interstellar absorption) corresponds to
$M_V\simeq-4.8$ which is marginally consistent with a relatively late
O star or a late B Iab supergiant (cf Wegner 2006).  As there are
indications that the spectrum is a composite spectrum of two early
type stars, it is possible that the LGGS magnitude is not of a single
star, therefore the magnitude may be overestimated.  However, the most
probable interpretation of the data is that of an OB supergiant of the
Iab type, showing no H$\alpha$ emission. The X-ray characteristics of
the source are consistent with a HMXB, with an X-ray flux of $\simeq
8-9\times10^{35}$ erg s$^{-1}$.

This object is recorded by \citet{Vilardell2006} as a low amplitude
variable in M31, M31V\_J00425892+4137271 (V=19.546, B=20.742), with a
period of 87.78days, but with no further classification.

\subsubsection{LGGS\_J004259.07+413731.4 -- Chance Superposition}

\begin{figure}
\centerline{\psfig{file=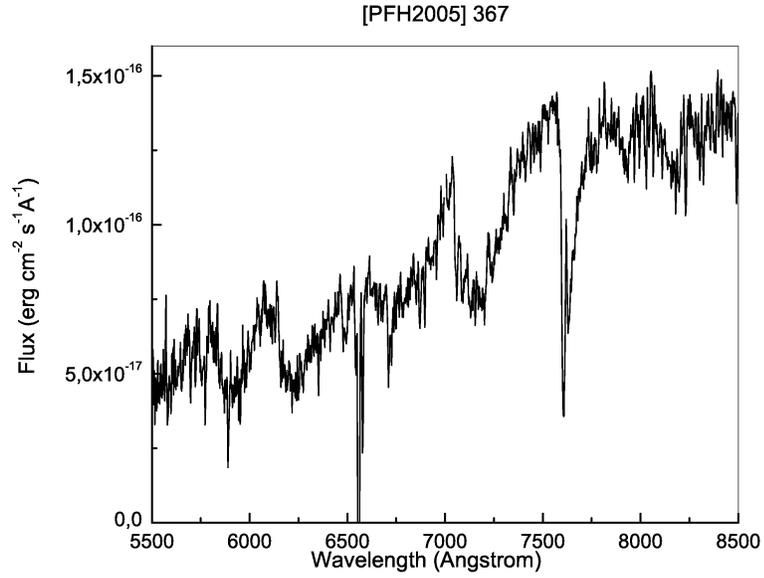,width=4.5in,angle=0}}
   \caption[]{ Part of the red spectrum of an M foreground star
   within the error circle of M31\_367.}
\label{367bspec}
\end{figure}

This object has the spectrum of an M5 foreground star (see
 Figure~\ref{367bspec}), and shows no significant variability
 \citep{Vilardell2006}. It must be noted that the colours recorded in
 LGGS for this star are not consistent with the spectral type of the
 star, and would correspond to an earlier spectral type. The spectral
 classification however overwrites the photometric information from
 LGGS.  While such stars can be X-ray-producing systems,
 \citep[e.g. in symbiotic X-ray binaries][]{Corbet2008}, they would be
 expected to have a much softer X-ray spectrum than observed.

In conclusion, the early type Iab supergiant LGGS\_J004258.94+413727.3
is the most probable optical counterpart of the X-ray source, and
therefore the object is a probable HMXB (SG-XRB).

\subsection{[PFH2005] 405 -- M31 LMXB in a Star Cluster}

There is one optical counterpart in the LGGS catalogue within the
3-sigma error circle of the X-ray source position,
LGGS\_J004309.86+411900.9.  There is a confirmed (unresolved) globular
cluster, SK059A (10.791125, 41.31700) registered at this position in
the Revised Bologna catalogue of M31 globular clusters, first recorded
in the survey of \citet{Kim2007}. However, the {\em optical} flux of
the object is known to be variable \citep[R-band period of 434$\pm$4
days and I-band period of 385$\pm$16d;][]{fliri2006}.  Fliri et
al. classify it as a semi-regular red variable.

The source correlates also with an unresolved radio source 37W153
\citep[S1412MHz=1.1$\pm$0.2mJy][]{walterbos1985}.  Finally, the object
has also been classified as an HII region by \citet{Caldwell2009} and
\citet{peacock2010}.

In X-rays, the source is transient
\citep{williams2004a,williams2006}. Its variability has been studied
by \citet{hofmann2013}, who classify the X-ray lightcurve as highly
variable.

\begin{figure}
\centerline{\psfig{file=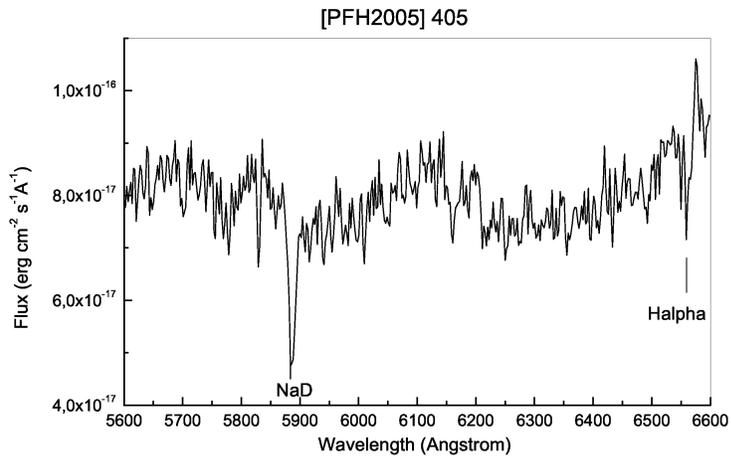,width=4.5in,angle=0}}
   \caption[]{ Part of the red spectrum of the optical counterpart
   of M31\_405, consistent with a K supergiant in M31.}
\label{405spec}
\end{figure}

The optical spectrum we have obtained is consistent with a relatively
late type star, and resembles a K supergiant (K2Iab), with significant
H$\alpha$ absorption \citep[as expected for early K supergiants, see
e.g.][]{Mallik1986}.  The spectrum (see Figure~\ref{405spec}) is very
similar for example to that of the K3Iab supergiant BN Car. The
spectrum allows the possibility of a globular cluster, but is
inconsistent with this hypothesis in two ways: there is H$\alpha$
absorption, and there is indication of emission lines in [OIII]
(5007\AA) and [NII] (6584\AA). The spectrum is blue-shifted by
$\simeq$180 km~s$^{-1}$ consistent with M31 membership.

The LGGS colours are entirely consistent (comparison with standard
values from Allen's Astrophysical quantities 4th edition) with an
early K giant and marginally consistent with a K0 (or late G)
supergiant, while the absolute magnitude at $M_V\simeq-6.3$ is only
consistent with an early K or late G supergiant. The infrared colours
available from 2MASS are compatible with a K5Iab supergiant (or a bit
later type).

The long period optical variability, radio emission, optical spectrum
and optical and infrared colours therefore strongly suggest that this
is a red long-period variable supergiant star.

The X ray source is highly variable and relatively hard (not as hard
as BeXRBs). If it is connected to the globular cluster then it is a
LMXB. If it is connected with the massive red supergiant, then it
would be an accreting binary, as single late type supergiants can emit
X-rays (e.g. FK Comae stars) only at much lower luminosities
($\gap$10$^{32}$ erg~s$^{-1}$ .  Given the late spectral type, the
object cannot be a HMXB (as the massive star in these cases has
spectral type O or B). So the object may belong to the poorly
populated class of symbiotic X-ray binaries, like XTE J1743-363
\citep{Smith2012}.  These types of objects can have X-ray luminosities
from few $10^{32}$ to 10$^{37}$ erg~s$^{-1}$ \citep{postnov2010}.

As the presence of a globular cluster is rather certain, with the
source being definitely extended (and not an HII region as no
characteristic lines were seen), it is also possible that there is a
bright long period variable coinciding in location with the
cluster. In this case the source is most probably connected with the
GC, and it is a LMXB.  If the radio emission is also connected to the
same source it could point to a microquasar.

\subsection{[PFH2005] 407 -- Strong M31 BeXRB Candidate}

 There are two counterparts in the LGGS catalogue within the 3-sigma
 error circle of the X-ray source, LGGS\_J004310.5+413852.0 and
 LGGS\_J004310.43+413854.7. We obtained a spectrum of the former,
 which is the brightest of the two, at V=19.96~mag.  The second
 candidate counterpart is a faint blue object with $V=21.39$

\begin{figure}
\centerline{\psfig{file=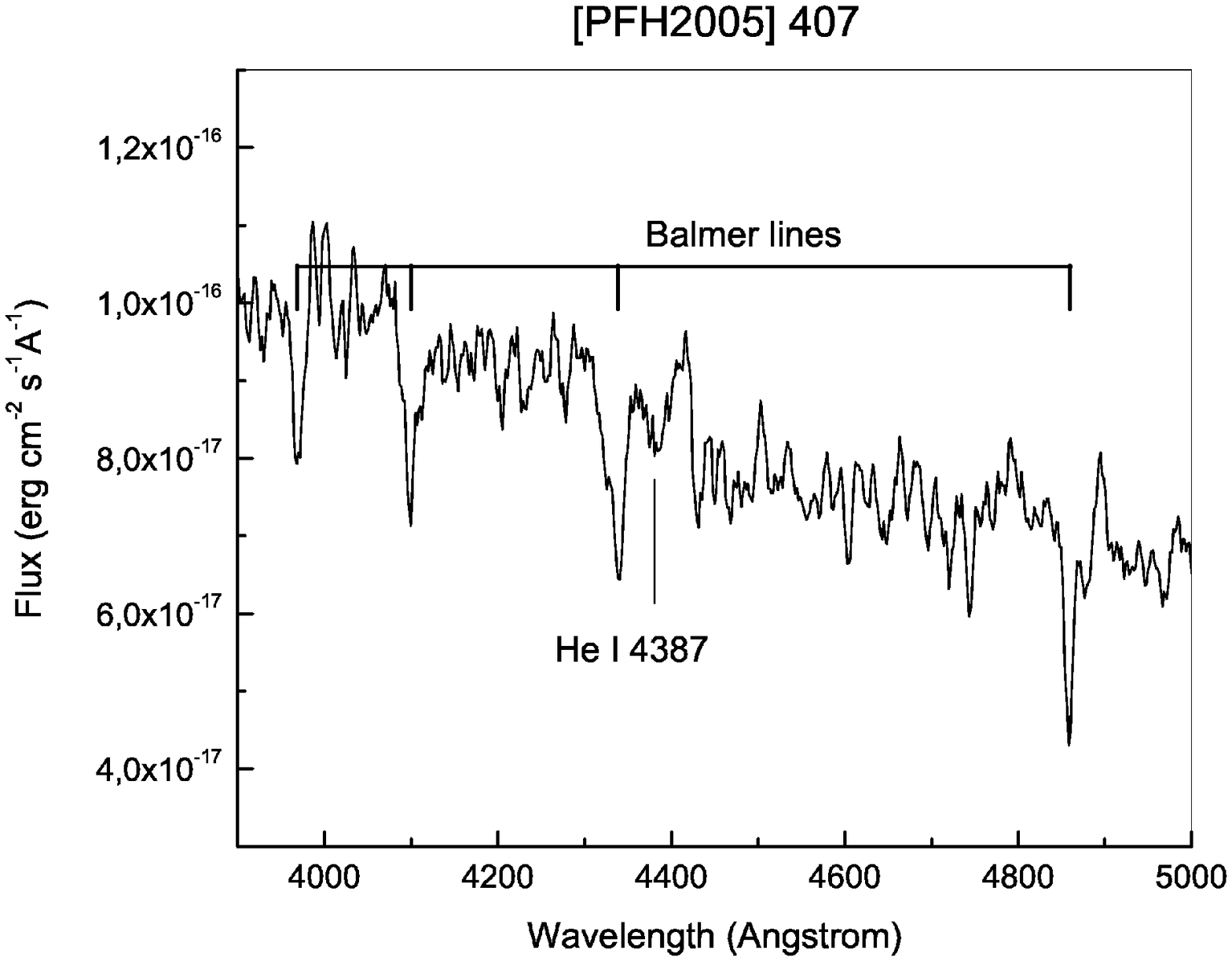,width=4.5in,angle=0}}
   \caption[]{ Blue part of the spectrum of the brightest optical counterpart
   of M31\_407, consistent with a B star in M31.}
\label{407bluespec}
\end{figure}

\begin{figure}
\centerline{\psfig{file=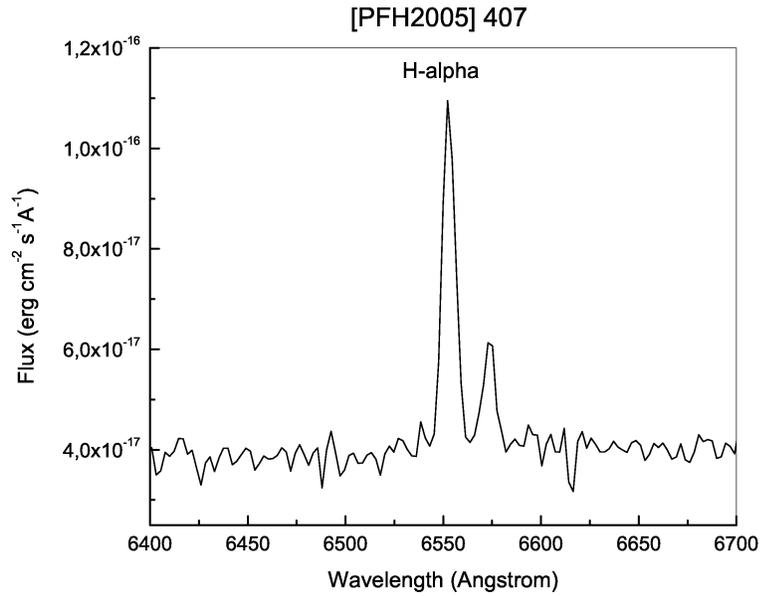,width=4.5in,angle=0}}
   \caption[]{ Red part of the spectrum of the brightest optical counterpart
   of M31\_407, showing H$\alpha$ emission.}
\label{407redspec}
\end{figure}

The spectrum of LGGS\_J004310.5+413852.0 is shown in
Figure~\ref{407bluespec} and in Figure~\ref{407redspec}. It shows
characteristic Balmer and HeI absorption lines at a velocity of
$-100$~km~s$^{-1}$.  Cross-correlation is good with VOR+59 97 (B2 V)
and a velocity of -140$\pm$100 km~s$^{-1}$, in agreement with our
qualitative assessment.

The HeI lines clearly classify the star as a B star. The magnitude,
colours and reddening free Q-parameter from LGGS (see
Table~\ref{Optical}) are entirely consistent with an early B star in
M31.  In the red part of the spectrum, there is significant
H${\alpha}$ emission, with a FWHM of 7.2\AA (after correcting for the
spectral resolution, as for source 235 above).  The star is a known
emission-line star \citep{Massey2007} with Ha magnitude of 19.72.  Due
to its proximity to an HII region \citep[also recorded as a SNR
candidate 89A;][]{braun1993} just outside the error circle \citep[HII
region ID2013;][]{Azimlu2011}, the spectrum may be contaminated by
diffuse emission, however, the H$\alpha$ image of the area shows no
diffuse emission at the location of the star. Therefore the H$\alpha$
emission seems to be related to the B star (but some contamination is
also possibly present).  Interestingly, the star is reported as
showing significant variability in the optical, with a period of
15.802446d \citep{Vilardell2006}.  In X-rays the source is classified
as hard and the hardness ratios are consistent with BeXRBs (see
discussion).  The X-ray luminosity of the source is $\simeq
3.6\times10^{35}$ erg~s$^{-1}$.

Therefore, the Be stellar counterpart and X-ray properties are
entirely consistent with a BeXRB. This is the first BeXRB identified
with some certainty in M31.

\subsection{[PFH2005] 466 -- M31 Supergiant HMXB candidate}

There are two possible counterparts in the LGGS catalogue within the
X-ray position error circle. The brightest and bluest,
LGGS\_J004333.64+411404.8, was the one observed.  The other object,
LGGS\_J004333.59+411407.7, is two magnitudes fainter and redder.

\begin{figure}
\centerline{\psfig{file=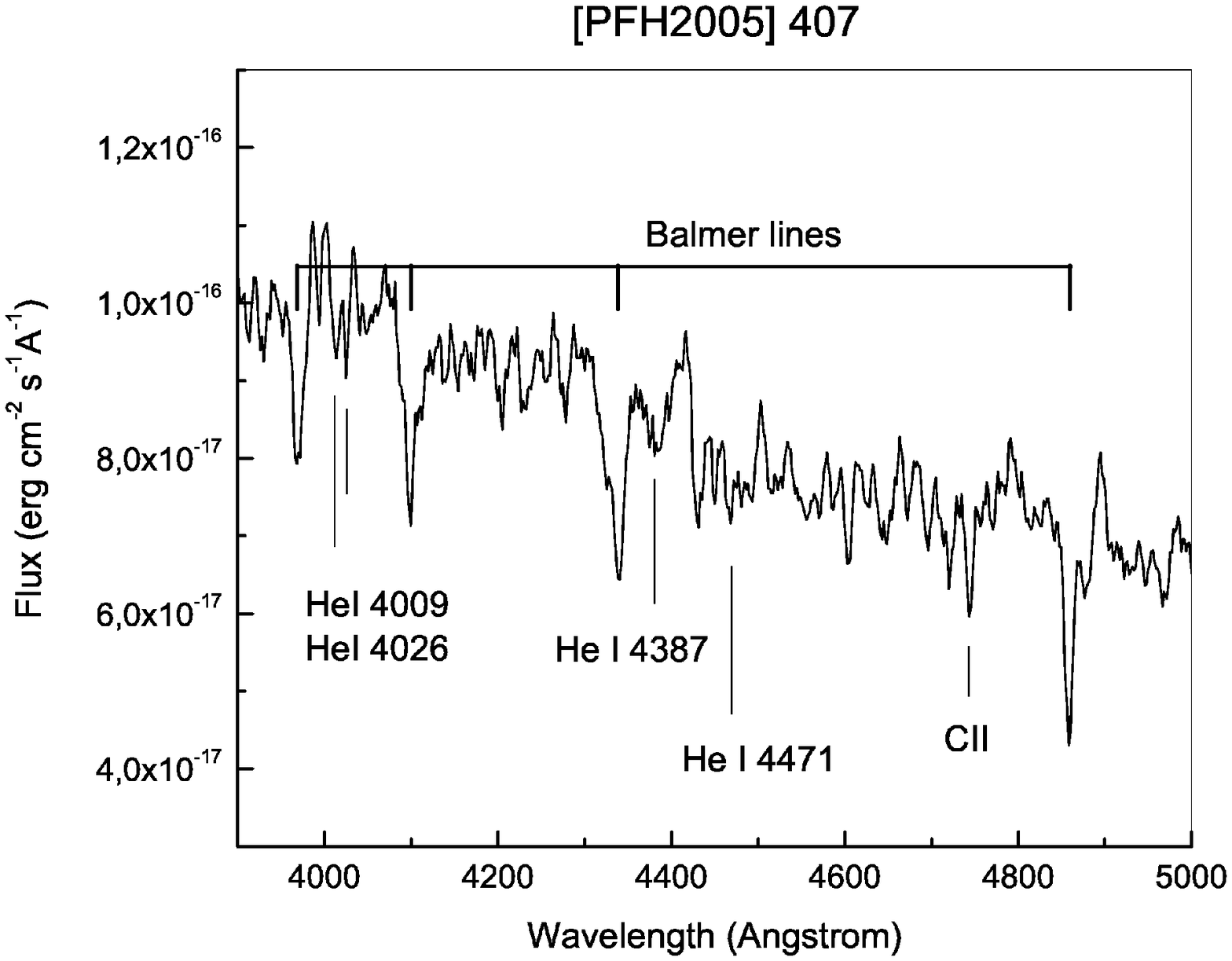,width=4.5in,angle=0}}
   \caption[]{ Blue part of the spectrum of the brightest optical counterpart
   of M31\_466, consistent with an O star in M31.}
\label{466bluespec}
\end{figure}

\begin{figure}
\centerline{\psfig{file=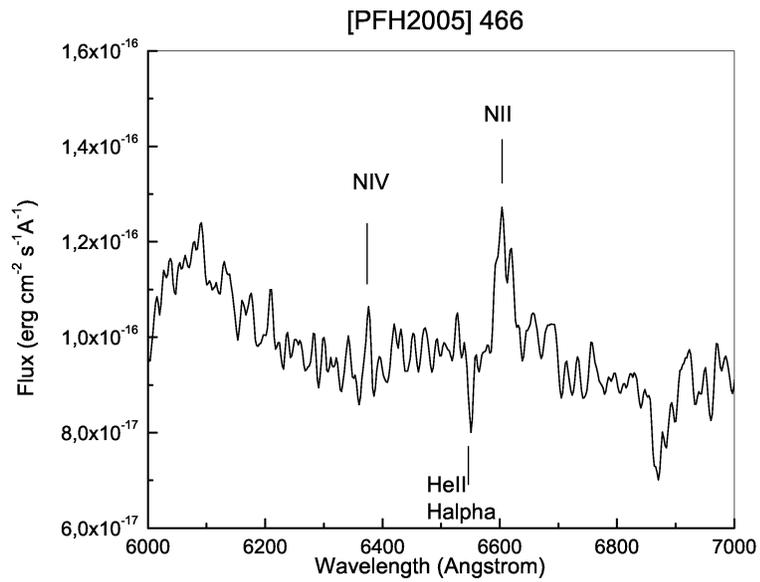,width=4.5in,angle=0}}
   \caption[]{ Red part of the spectrum of the brightest optical counterpart
   of M31\_466, showing clear NII emission, pointing to a late type O supergiant.}
\label{466redspec}
\end{figure}

The spectrum is blue-shifted by about $-$200 km~s$^{-1}$, confirming
M31 membership.  The spectrum shows a correlation peak with VOR+59 97
(B2 V) with a velocity of -130$\pm$80 km~s$^{-1}$. The blue part of
the spectrum of LGGS\_J004333.64+411404.8 (Figure~\ref{466bluespec})
shows HeI (4009 and 4471) and possibly SiII 4028-30 absorption lines
consistent with a B star.  The red part of the spectrum
(Figure~\ref{466redspec}) shows strong NII$\lambda6610$ emission,
normally seen in later type O supergiants \citep{Walborn2001}.  For
such a spectral type one would also expect HeII lines, which are
probably too weak for the S/N ratio of our spectra.  Certainly the
presence of HeI rules out any spectral type earlier than O9.
According to the Atlas of \citet{WalbornandFitzpatrick1990}, the
spectral characteristics of LGGS\_J004333.64+411404.8 may also be
consistent with ON9.7Ia+.

The optical magnitude and colours, as well as the value of the
reddening free Q-parameter of LGGS\_J004333.64+411404.8 (see
Table~\ref{Optical}) are more consistent with a relatively heavily
reddened B-type supergiant. The Q-parameter value of $-0.69$ points
toward an early B star, between B1 and B3 depending on the luminosity
class. Given the V magnitude of 19.49 and estimating the reddening
from the observed $B-V=0.544$ and the estimated $B-V\simeq-0.2$ to
$-0.1$, according to the Q-parameter value, we estimate an absolute
magnitude around -6.5 consistent with an O9 supergiant.  In
conclusion, we classify the object as an OB supergiant.

Young O stars are strong, hard, and variable X-ray sources that often
dominate the X-ray emission from young clusters, but their X-ray
luminosities only extend up to L$_{\rm X}$ = 10$^{34}$ erg s$^{-1}$.
Therefore, because of the high X-ray luminosity
(${\sim}5{\times}10^{35}$ erg s$^{-1}$ of [PFH2005] 466, the observed
X-rays are more likely from accretion onto a compact companion to this
O-supergiant, making this source a good HMXB candidate, of the SgXRB
type.

\subsection{[PFH2005] 581 -- Galactic CV}

There are two counterparts in the Massey catalogue, the brightest and
bluest is LGGS\_J004445.06+415153.4, which we observed, and a fainter
also quite blue object LGGS\_J004444.97+415151.3.

\begin{figure}
\centerline{\psfig{file=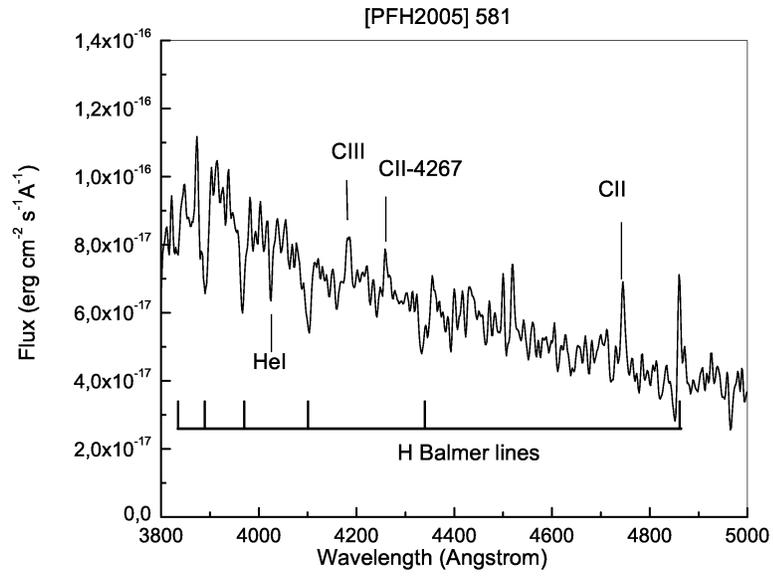,width=4.5in,angle=0}}
   \caption[]{ Blue part of the spectrum of the brightest optical counterpart
   of M31\_581, consistent with a Galactic cataclysmic variable star.}
\label{581bluespec}
\end{figure}

The optical spectrum of LGGS\_J004445.06+415153.4
(Figure~\ref{581bluespec}) shows significant emission lines of CII
(4267, 4738, 4745\AA), as well as H-beta in emission (H-gamma seems to
be filled-in by emission; earlier Balmer lines seen in absorption) and
weak CIII emission (4187\AA). There is also clear absorption of
HeI4026. Such a spectrum is reminiscent of a cataclysmic variable
(dwarf nova) in eruption \citep[e.g.,][]{Joergens2000}, although it
lacks the expected HeI and HeII emission lines that should also be
present in such spectra.  The blue colour, characteristic emission
lines, X-ray emission and relatively low velocity ($\simeq$50
km~s$^{-1}$, are consistent with a foreground cataclysmic variable.
As the X-ray spectrum is hard, the object is most probably an
intermediate polar with expected L$_{\rm X}\simeq10^{31}-10^{33}$
erg~s$^{-1}$ (which is also consistent with the observed X-ray flux of
$1.8{\times}10^{-14}$ mW~m$^{-2}$, since the source is within the
Galaxy).

\subsection{[PFH2005] 599 -- Chance Superposition}

\begin{figure}
\centerline{\psfig{file=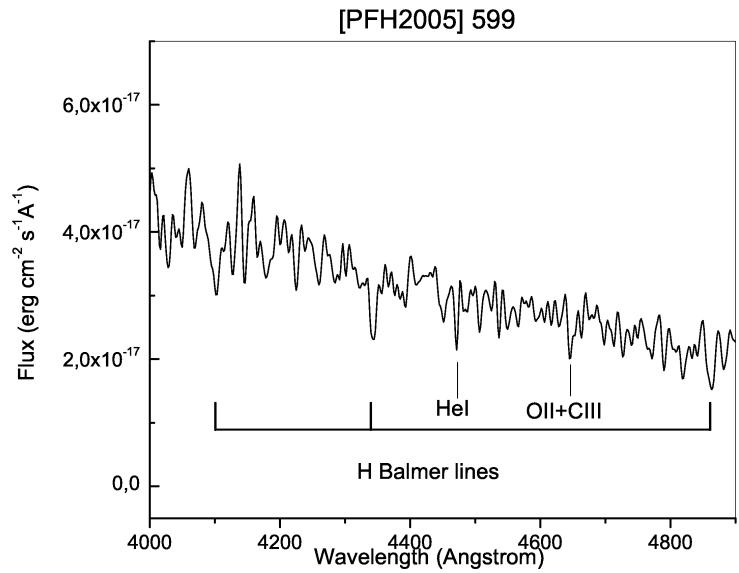,width=4.5in,angle=0}}
   \caption[]{ Blue part of the spectrum of the brightest optical counterpart
   of M31\_599, consistent with a Galactic early type star.}
\label{599bluespec}
\end{figure}

There is only one optical counterpart candidate listed in LGGS,
LGGS\_J004456.78+413548.0.  The spectrum of this faint blue
counterpart candidate (Figure~\ref{599bluespec}) shows weak Balmer and
HeI absorption lines, characteristic of an early B star, and no
emission lines.  Its radial velocity is red-shifted by $\simeq$80
km~s$^{-1}$, showing that it is a foreground object.  The X-ray source
is hard and weak. The optical spectrum is not consistent with a
cataclysmic variable. Therefore, most probably the object observed is
not the correct counterpart. Within the X-ray position error circle
(3$\sigma$; 5.26$''$) there are several faint sources visible in the
LGGS images, although they are not catalogued.  \citet{Vilardell2006}
list 7 faint sources in addition to LGGS\_J004456.78+413548.0, within
the error circle. They range in magnitude between $V=21.8$ and
$V=24.8$~mag. The brightest of the seven is a blue variable star,
M31V\_J00445707+4135466, with a period of 80d.  Further observations
are needed to identify the optical counterpart of the X-ray source
[PFH2005]599.

\subsection{[PFH2005] 612 -- M31 HMXB candidate}

\begin{figure}
\centerline{\psfig{file=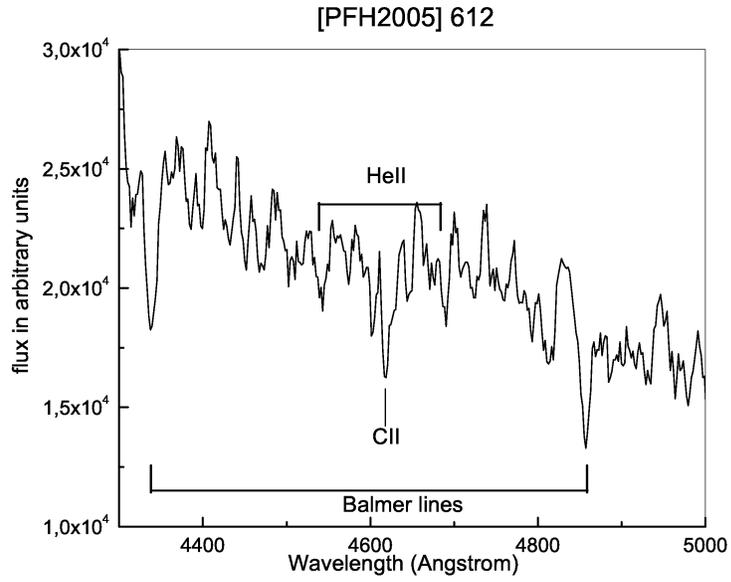,width=4.5in,angle=0}}
  \caption[]{ Part of the blue spectrum of [PFH2005] 612, characteristic of an early type star }
\label{612bluespec}
\end{figure}

There is a single source catalogued in LGGS within the 3-sigma error
circle of the X-ray source, object LGGS\_J004506.46+420615.8. The
spectrum obtained for the star suffers from relatively low S/N
(Figure~\ref{612bluespec}), however, it does show a strong blue
continuum with Balmer and HeII lines as well as possible CII in
absorption, blueshifted by about $-$150~km~s$^{-1}$, which is
consistent with M31 membership. The spectrum is consistent with a
mid-to late O dwarf star in M31.  Cross-correlation peaks with B+57 18
(B3 IV) with a velocity of $-$110$\pm$160 km s$^{-1}$, so reddening
may be affecting the correlation, but in any case, it is an early-type
star in M31.

In the red part of the spectrum H$\alpha$ is not resolved.  It is either
filled in by emission and/or weak and blended with adjacent HeII lines
(6527\AA, 6560\AA).

The optical magnitude and colours, as well as the value of the
reddening free Q-parameter of LGGS\_J004506.46+420615.8 (see
Table~\ref{Optical}) are also consistent with a O5-O9 main sequence
star in M31.

The X-ray properties of the source indicate a hard spectrum and a
luminosity of L$_{\rm X}$=5.2$\times$10$^{35}$ erg~s$^{-1}$.

The early spectral type (late O) of the star and its X-ray properties
are consistent with a HMXB, however the lack of significant H$\alpha$
emission weakens the identification of the source with a BeXRB.  It
must also be noted that although the star appears to be a single
counterpart in the LGGS catalogue, inspection of the images shows other
possible fainter counterparts within the error circle, some of which
are particularly visible in the $I$ image.

Overall, the source is an HMXB candidate, but not one of our
strongest.

\subsection{[PFH2005] 620 -- M31 HMXB Candidate}

\begin{figure}
\centerline{\psfig{file=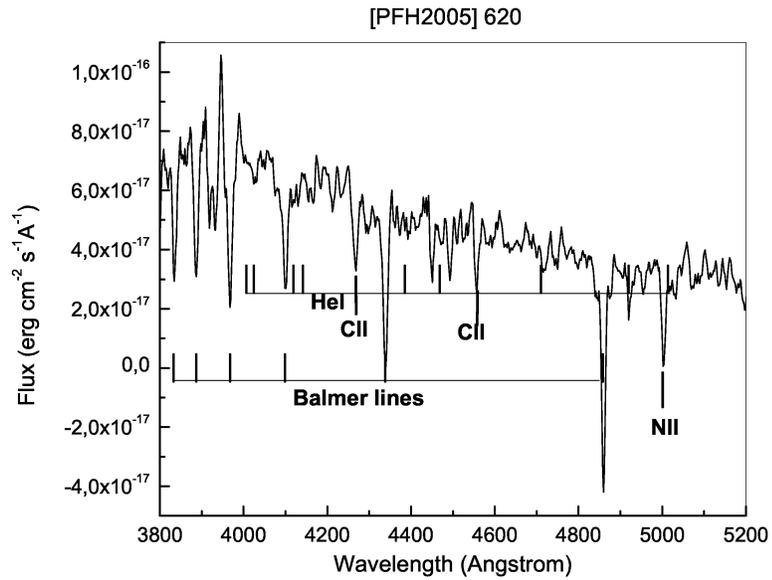,width=4.5in,angle=0}}
  \caption[]{ Part of the blue spectrum of [PFH2005] 620, showing an early type star. }
\label{620bluespec}
\end{figure}

Within the error circle of the X-ray position (3sigma 4.26'') there is
one object recorded in LGGS, LGGS\_J004513.59+413805.7, although in
the near infrared image at least another five very faint sources are
visible. Indeed, \citet{Vilardell2006} report another 8 faint objects
within the 3sigma error circle of the X-ray source position with V
magnitudes between 22.35 and 24.77~mag.  The spectrum of the single
LGGS counterpart is shown in Figure~\ref{620bluespec}.  The star has a
blue continuum with Balmer lines in absorption, no later than late B
(from the appearance of the Balmer lines). The S/N does not allow
accurate classification of the spectrum, however there are indications
of some HeI and CII absorption lines. No emission lines are seen in
the spectrum. The H$\alpha$ line is also seen in absorption. The
velocity of the star (${\simeq}-$200 km~s$^{-1}$) is consistent with
M31 membership.  Cross-correlation yields plausible peaks with several
of our B-type standards, including VOR+59 97 (B2 V) with a velocity of
$-$160$\pm$110 km~s$^{-1}$.  The magnitude and colours given in LGGS
for the observed star (see Table~\ref{Optical}) do not provide a
consistent classification for the object: the reddening free Q
parameter value around 0 suggests a late B or early A star, while the
V magnitude is too bright for such a late type star (unless it is a
supergiant). It is possible that the photometry is contaminated by
nearby faint objects.  The X-ray properties of the source suggest a
hard spectrum and a luminosity of $5.7\times10^{35}$ erg~s$^{-1}$.

Thus, this is still a HMXB candidate, but not one of the better
candidates due to the lack of emission lines, and relatively uncertain
spectral classification.

\subsection{[PFH2005] 705 -- Chance Superposition}

There is a single counterpart in the LGGS catalogue for this object,
LGGS\_J004604.83+415142.4.  The spectrum shows a F-G foreground star
with no H$\alpha$ emission. The X-ray source is classified as hard, on
the basis of its hardness ratio values, therefore the star observed is
not the correct counterpart.  On the LGGS images there are more than
three faint red objects within the 3-sigma error circle, which are
candidate counterparts, which however are not recorded in
\citet{massey2006}.

\subsection{[Stiele2011] 1716 -- M31 HMXB Candidate}

There is a single counterpart candidate in the LGGS catalogue within the
X-ray error circle, LGGS\_J004557.04+414830.0, with $V=20.017$ and
$B-V=0.123$.  Except for this relatively bright blue object, there
appears to be another object within the error circle, visible in the
near infrared image, but not recorded in the LGGS catalogue. This very
red object is also recorded in 2MASS (2MASS J00455703+4148322) and in
the WISE All-Sky data release catalogue \citep[WISE
J004556.96+414832.3,][]{Cutri2012}.

\begin{figure}
\centerline{\psfig{file=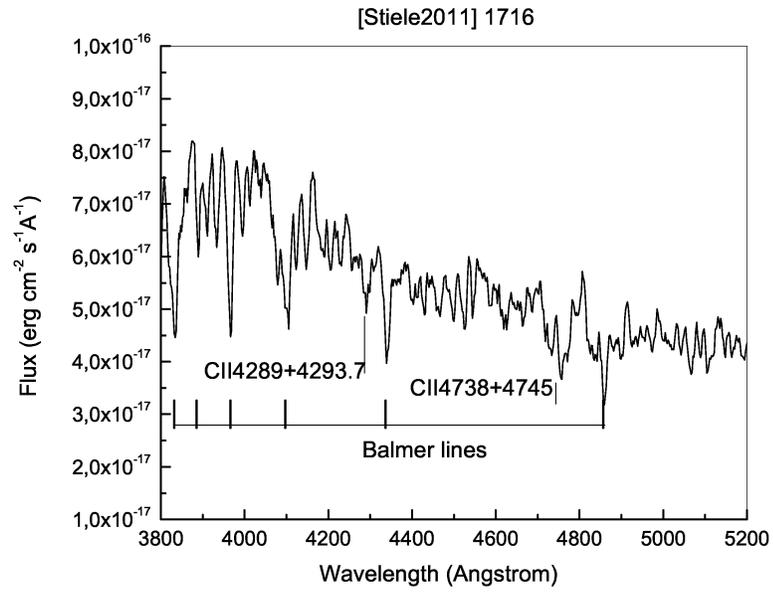,width=4.5in,angle=0}}
  \caption[]{ Part of the blue spectrum of [Stiele2011] M31\_1716, characteristic of a Be-type star. }
\label{1716spec1}
\end{figure}

\begin{figure}
\centerline{\psfig{file=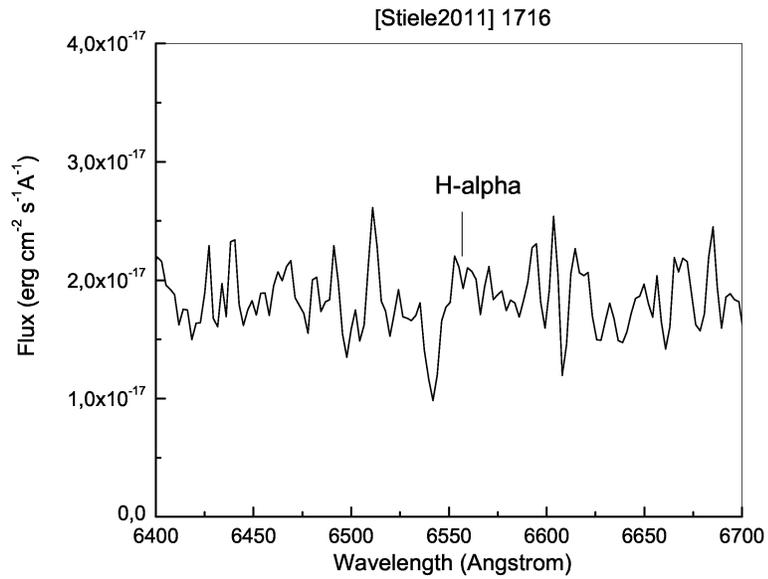,width=4.5in,angle=0}}
  \caption[]{ Part of the red spectrum of [Stiele2011] M31\_1716, showing the H$\alpha$ line. }
\label{1716spec2}
\end{figure}

We have obtained a spectrum for the bright blue counterpart
LGGS\_J004557.04+414830.0.  The blue part of the spectrum (see
Figure~\ref{1716spec1}) reveals a hot star with Balmer line
absorption, and two CII line pairs that can be identified with
relative certainty, CII4738+4745, CII4289+4293.7, which are common
features in the spectra of Be stars.  The Balmer lines from
H$\epsilon$ to to H$\beta$ are clearly seen in absorption at a
velocity of $\simeq$-260 km~s$^{-1}$, confirming M31
membership. Cross-correlation yields peaks with VOR+59 97 (B2 V) at a
velocity of $-$170$\pm$160.  The red part of the spectrum (see
Figure~\ref{1716spec2}) shows the H$\alpha$ line almost completely
filled in with emission, resulting in a very weak line. This is a
known feature of Be stars.  To classify the star as a Be star, the
H$\alpha$ line should be seen in emission. In this case, the line is
almost completely filled in, which indirectly indicates
emission. Therefore, the object is likely a Be star.  The hardness
ratios are completely consistent with the SMC BeXRBs. The X-ray flux
corresponds to $5.1\times10^{36}$ erg~s$^{-1}$.  Therefore, the object
is a good candidate BeXRB in M31.

\section{Discussion and Conclusions}\label{discussion}

Table~\ref{Classifications} summarizes our classifications, on the
basis of the spectral characteristics of the candidate optical
counterparts. In all but two cases we have identified a source that
could provide the observed X-ray emission. Among these objects, there
are 8 that are possible (5) or probable (3) high mass XRB in M31.  We
provide their radial velocities, along with the velocity of the M31
gas disc at their projected location \citep[as determined from the
  21~cm maps of][]{chemin2009} in Table~\ref{velocities}.  The
velocities are in general agreement with ours, which have relatively
low $\sim$100 km~s$^{-1}$ precision.  The largest outlier is [PMH 235],
which is 200 km~s$^{-1}$ different from the local gas, but still not
enough different to suggest that it is a high velocity star.

\begin{figure}
\centerline{\psfig{file=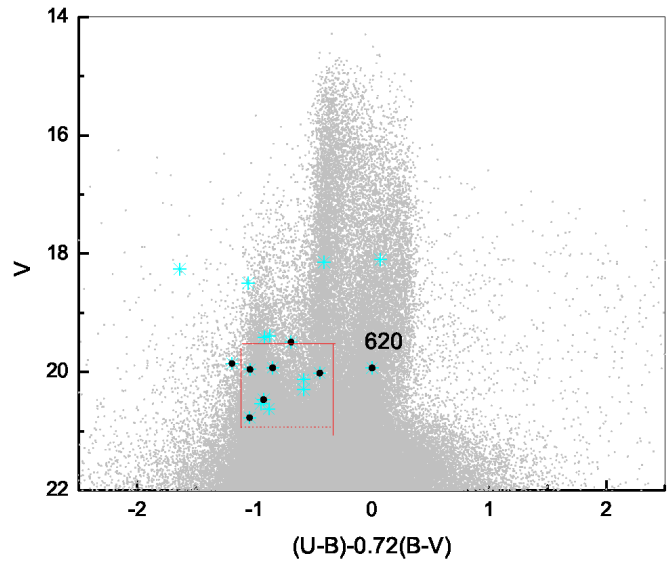,width=4.5in,angle=0}}
  \caption[]{Colour-magnitude diagram of apparent magnitude vs. the
reddening free Q-parameter Q=$U-B-0.72(B-V)$, for all stars in the
LGGS-M31 catalogue with both U-B and B-V colours measured (grey
dots). Cyan bullets mark the location of all objects in the present
study (according to Table~\ref{Optical}). Black dots signify the
subset of objects for which we have given a classification of a
candidate HMXB of class A or B.  The red box indicates the locus of
BeXRBs in the SMC and the Galaxy moved to the distance of M31 (the
faint limit is the observational limiting magnitude of our survey).}
\label{cmd}
\end{figure}

It is interesting to examine the location of these objects in
characteristic diagrams in the optical and in the X-rays, in
comparison to the loci of confirmed HMXBs of other galaxies on the
same diagrams.  Figure~\ref{cmd} shows a diagram of apparent magnitude
versus the reddening free Q-parameter Q$=U-B-0.72(B-V)$, for all stars
in the LGGS-M31 catalogue with both U-B and B-V colours measured (grey
dots). The Q parameter is used instead of the customary B-V colour, in
order to compensate for the variable reddening to the individual
sources of interest here.  Clearly, all but one of the HMXB candidates
lie within the expected locus. Object 620 lies outside the box,
however the spectral class is consistent with an early type star,
suggesting a problem with the LGGS colours for this star.

\begin{figure}
\centerline{\psfig{file=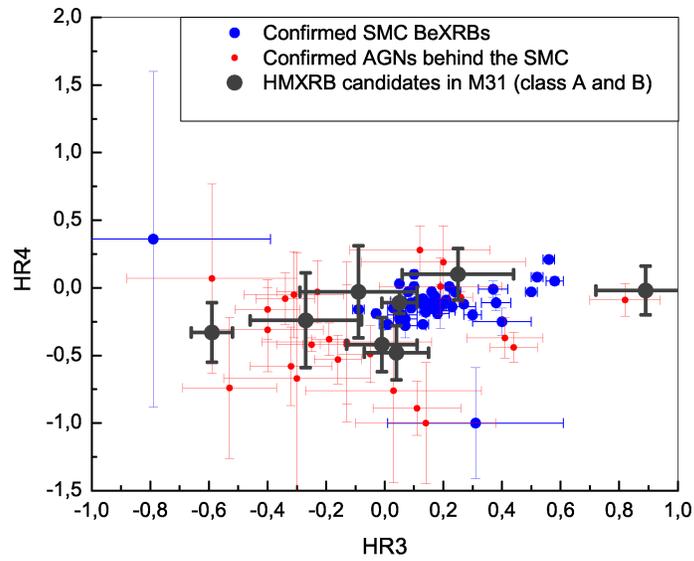,width=4.5in,angle=0}}
\caption[]{X-ray hardness ratio diagram, of HR4 versus HR3 (see
Table~\ref{X-ray}.  Blue dots mark confirmed BeXRBs in the Small
Magellanic Cloud (SMC).  Red dots mark confirmed AGNs behind the SMC,
from \citep{sturm2013}. Grey shows class A and class B HMXB candidates
from Table~\ref{Classifications} (see Section~\ref{discussion}).}
\label{HR}
\end{figure}

Figure~\ref{HR} shows a X-ray hardness ratio diagram, of HR4 versus
HR3 (see Table~\ref{X-ray}.  With blue dots we mark the location of
confirmed BeXRBs in the Small Magellanic Cloud (SMC), while with red
dots of confirmed AGNs behind the SMC, from \citep{sturm2013}. In grey
we show the location of class A and class B HMXB candidates from
Table~\ref{Classifications}. Generally, BeXRBs and AGNs are relatively
well separated on this diagram in the SMC, although there is
overlap. Our candidates lie within the locus of the SMC BeXRBs, taking
into account the observational errors, although on average they are
somewhat softer that the SMC objects. Indeed, our only BeXRB candidate
(407) lies within the SMC locus, and so does 1716st which is
classified as a probable HMXB, but as mentioned above the absence of
H$\alpha$ absorption probably indicates emission that fills in the
absorption line, which would make the object a very good BeXRB
candidate.

In conclusion, we have observed, reduced, and classified 20 spectra of
bright blue optical counterpart candidates of 17 X-ray sources in the
M31 field.  We have found 3 strong HMXB candidates, and 5 acceptable
candidates in the field of M31. These are likely only a fraction of
the HMXBs in M31 when compared to expectations based on the star
formation rate and comparisons with the Galaxy.  Significantly more
could be found with improved data sets.  For example, precise source
locations could be obtained by {\it Chandra} imaging and matched with
deeper and better-resolved stellar photometry, such as that of the
Panchromatic Hubble Andromeda Treasury \citep[][Williams et al., in
prep.]{dalcanton2012}.

While our candidates require very high-quality spectral follow-up on
larger telescopes in order to look for radial velocity variations that
would confirm their HMXB nature, the relative numbers, optical
spectra, optical photometry, and X-ray properties of our best HMXB
candidates suggests they should be the focus of dedicated high
signal-to-noise studies at higher spectral resolution in the future.

Support for this work was provided by NASA under grant NNX06AF58G.

\begin{landscape}

\begin{table*}
 \begin{center}\scriptsize
  \caption{Log of stellar objects observed in the direction of M31.}
  \label{Log}
  \begin{tabular}{ccccccccc}
   \hline\hline\noalign{\smallskip}
   \multicolumn{7}{c}{{M31}}\\
   \hline\noalign{\smallskip} { Optical ID}$^\spadesuit$ &
   {xID}$^\clubsuit$ &  {RA} & {Dec} & {Offset} & {Offset} & {Date} &
   {Band}$^\heartsuit$ & {Exposure}\\
   {(Massey)} & {(Piestch)} & {(J2000)} & {(J2000)} & arcsec &
   $\sigma$ & {(yyyy-mm-dd)} & (B,R) & {(s per Band)} \\
   \noalign{\smallskip} \hline\noalign{\smallskip}
LGGS\_J004123.75+411459.6  & \textbf{136}  & 00 41 23.75 & 41 14 59.6
& 1.5 & 1.0 &  2008-10-26 & B,R & 3x2700\\
LGGS\_J004130.37+410500.9  & \textbf{146}  & 00 41 30.37 & 41 05 00.9
& 2.5 & 3.1 &  2006-10-23 & B,R & 4x900\\
LGGS\_J004210.24+405149.4 & \textbf{224a}  & 00 42 10.24 & 40 51 49.4
&  2.6 & 3.3 &  2007-10-10 & B,R & 3x2700\\
LGGS\_J004210.38+405148.1  & \textbf{224b}  & 00 42 10.38 & 40 51 48.1
& 0.7 & 0.9 & 2009-10-18 & B,R& 3x1800\\
LGGS\_J003937.29+404925.8  & \textbf{235-st} & 00 39 37.3  & 40 49
25.8 & 4.8 & 2.6 &  2011-09-26 & B,R & 4x2700\\
LGGS\_J003937.27+404928.9  & \textbf{235-st} & 00 39 37.3  & 40 49 28.9 & 2.6 & 1.4 & 2012-11-08 & B,R & 3x2400\\
LGGS\_J004216.78+404814.4 & \textbf{242}  & 00 42 16.78 & 40 48 14.4 &
2.3 & 3.1 &2006-10-26 & B,R & 950,7x1200\\
       &               &             &     & &        &  2006-10-29 & B,R & 3x1200\\
LGGS\_J004218.72+411401.2 & \textbf{246}  & 00 42 18.72 & 41 14  1.2 &
0.2 & 0.1 & 2006-11-16 & B   & 1200,2x900\\
LGGS\_J004231.23+410435.3 & \textbf{278}  & 00 42 31.23 & 41  4 35.3 &
1.3 & 1.9 & 2007-08-11 & B   & 5x1800\\
LGGS\_J004259.07+413731.4 & \textbf{367}  & 00 42 59.07 & 41 37 31.4 &
2.6 & 3.7 & 2008-09-23 & B,R & 1800,2x2700\\
LGGS\_J004258.94+413727.3 &\textbf{367obj1} & 00 42 58.94& 41 37 27.3&
2.3 & 3.3 & 2009-10-18 & B,R & 3x1800\\
LGGS\_J004309.86+411900.9 & \textbf{405}  & 00 43  9.86 & 41 19  0.9 &
0.2 & 0.1 & 2009-07-24 & B,R & 3x1800\\
LGGS\_J004310.50+413852.0 & \textbf{407}  & 00 43 10.50 & 41 38 52.0 &
2.0 & 1.6 & 2008-09-26 & B,R   & 5x1800\\
LGGS\_J004333.64+411404.8 & \textbf{466}  & 00 43 33.64 & 41 14 04.8 &
3.1 & 1.6 & 2007-10-09 & B,R & 2700,2x2500\\
LGGS\_J004445.06+415153.4 & \textbf{581}  & 00 44 45.06 & 41 51 53.4 &
1.4 & 0.8 & 2006-11-14 & B   & 3x2700\\
LGGS\_J004456.78+413548.0& \textbf{599}  & 00 44 56.78 & 41 35 48.0 &
0.1 & 0.1 & 2006-11-14 & B   & 1800,3x2700\\
LGGS\_J004506.46+420615.8 & \textbf{612}  & 00 45 06.46 & 42 06 15.8 &
2.4 & 0.9 & 2006-11-16 & B   & 3x2700\\
LGGS\_J004513.59+413805.7 & \textbf{620}  & 00 45 13.59 & 41 38  5.7 &
2.8 & 1.5 & 2008-09-23 & B,R & 3x2700\\
LGGS\_J004604.83+415142.4 & \textbf{705}  & 00 46 04.83 & 41 51 42.4 &
3.0 & 2.0 & 2007-11-03 & B,R & 5x1800 \\
LGGS\_J004557.04+414830.0 & \textbf{1716-st} & 00 45 57.04 & 41 48 30
& 1.1 & 1.1 & 2012-20-08 & B,R & 3x2400\\
\noalign{\smallskip} \hline\noalign{\smallskip}
\noalign{\smallskip}
  \end{tabular}
 \end{center}
$^\spadesuit$ Optical source numbers from \citet{massey2006} \\
$^\clubsuit$ X-ray source numbers from PFH2005 (M31).\\
$^\heartsuit$ B = blue (3750 - 5400\AA), R = red (5200 - 8000\AA) \\
$^1$ the spectrum is a composite of Massey sources
LGGS\_J004353.35+411656.1,LGGS\_J004353.36+411656.1,LGGS\_J004353.37+411656.9,LGGS\_J004353.47+411656.8
(see text)
\end{table*}


\begin{table}
 \begin{center}\scriptsize
  \caption{X-ray properties of stellar spectra from Table~\ref{Log}.}
  \label{X-ray}
  \begin{tabular}{rrrr|rccccc|l}
   \hline\hline\noalign{\smallskip}
   \multicolumn{3}{c}{{Object Name}}&Detection Likelihood & \multicolumn{6}{c}{{X-ray Properties}} & {Class} \\
   \hline\noalign{\smallskip} {[PFH 2005]} & {[Stiele 2011]}&Chandra&LH& {f$_{\rm X}$} & {f$_{X}$/f$_{\rm Opt}$} & {HR1} & {HR2} & {HR3} & {HR4} & \\
    &  &&  &{(mW/m$^2$)}
    & &  &  &  &  & \\
   \noalign{\smallskip} \hline\noalign{\smallskip}
   & 235  &  &19.8 & 6.95(1.0) e-15 &  & -  & - &0.89(0.17) &-0.02(0.18) & $<$hard$>$ \\
   & 1716 &  & 460 & 2.84(0.2) e-14 &  &0.59(0.26) &0.71(0.07)&0.05(0.06) &-0.11(0.08) & $<$hard$>$\\
 \textbf{136}&623&     &83 & 1.3(0.2) e-14 &    & 0.11(0.12) &-0.36(0.14)  & -0.52(0.29) &  -        & $<$fg$>$\\
                        &&&& 1.03(0.22) e-14&   & 0.29(0.14) &-0.27(0.16) & -0.74(0.28) &-&\\
 \textbf{146}&-&     &9.3 & 2.6(0.8) e-15 &    & 0.97(0.18) & -0.26(0.22) & -0.25(0.34) &  0.46(0.30) & SNR    \\
 \textbf{224}&811 &     &7.5 & 2.4(0.8) e-15 &    &-0.53(0.25)  &     -      &  -        &   -       & $<$SSS$>$  \\
              &&&            & 2.78(0.78)e-15&    &-0.46(0.46)  &0.48(0.46) &-0.83(0.36)&  -& $<$hard$>$    \\
 \textbf{242}&842 &     &76 & 5.5(0.7) e-15 &    & 0.48(0.10) & -0.90(0.10) &           &           & $<$SNR$>$  \\
            &&&             & 6.79(0.7)e-15 &    & 0.54(0.09) & -0.55(0.10) &-1.00(0.15)& - &$<fg>$ \\
 \textbf{246}&855 &r3-44&54000 & 4.8(0.04) e-13 &    & 0.45(0.01) & 0.32(0.01)  & -0.17(0.01) & -0.39(0.01) & GlC   \\
               &&&             & 4.21(0.37)e-13 &    & 0.51(0.01) & 0.32(0.01)  & -0.17(0.01) &-0.36(0.01) & GlC\\
 \textbf{278}&925 &     &1000 & 3.9(0.1) e-14 &    & 0.31(0.04) & -0.01(0.04) & -0.57(0.04) & -0.26(0.13) & $<$hard$>$ \\
              &&&             & 2.49(0.15)e-14 &   & 0.23(0.06) & -0.02(0.06) & -0.59(0.07) &-0.33(0.22)& $<$hard$>$\\
 \textbf{367}&1097&     &86 & 1.3(0.2) e-14 &    &   -      & 0.70(0.18)  & -0.02(0.13) & -0.38(0.26) & $<$hard$>$\\
&&&                         & 1.17(0.14)e-14&    &0.86(0.36)&0.64(0.14)&0.04(0.11)&-0.48(0.20) & $<$hard$>$ \\
 \textbf{405}&1152&r3-16&14000 & 1.7(0.02)e-13 &    & 0.47(0.02) & 0.25(0.01)  & -0.29(0.01) & -0.39(02) & $<$XRB$>$\\
&&&                            & 1.44(0.22)e-13 &   & 0.49(0.02)&0.26(0.02)&-0.28(0.02)&-0.38(0.02)& $<$XRB$>$\\
 \textbf{407}&1156&     &28 & 5.3(0.9) e-15 &    &     -     & 0.85(0.33)  &  0.29(0.17) &  0.08(0.18) & $<$hard$>$\\
&&&                         &4.77(0.94)e-15 &    &     -&0.98(0.32)&0.25(0.19)&0.10(0.19)& $<$SNR$>$\\
 \textbf{466}&1257&     &36 & 5.9(0.8) e-15 &    &   -     & 0.59(0.17)  & -0.18(0.13) & -0.27(0.23) & $<$hard$>$ \\
&&&                         & 6.4(1.3) e-15 &    &   -     & 0.58(0.23)  & -0.27(0.19) &-0.24(0.35)& $<$hard$>$\\
&&&                            & 3.91(0.06)e-13 &    & 0.49(0.02) & 0.13(0.01) &-0.45(0.01)& -0.84(0.03)& $<$GlC$<$ \\
 \textbf{581}&1514&     &140   & 1.8(0.2) e-14 &    & 0.43(0.25) & 0.55(0.10)  & -0.16(0.09) & -0.50(0.18) & $<$hard$>$ \\
&&&                            & 9.1(0.9) e-15 &    & 0.57(0.19) & 0.49(0.10)  & -0.23(0.09) &-0.51(0.18)& $<$hard$>$  \\
 \textbf{599}& -  &     &9.9 & 3.5(0.9) e-15 &    & 0.02(37) &    -      & 0.40(0.31)  &  0.34(0.26) & $<$hard$>$ \\
 \textbf{612}&1579&     &16 & 7.7(2.0) e-15 &   &   -       & 0.73(0.23)  & -0.15(0.21) & -0.25(0.38) & $<$hard$>$ \\
&&&                         & 9.7(1.2) e-15 &   & 0.73(0.27)& 0.54(0.14)  & -0.01(0.12) & -0.42(0.20) & $<$hard$>$\\
   \noalign{\smallskip} \hline\noalign{\smallskip}
   \noalign{\smallskip}
  \end{tabular}
 \end{center}
\end{table}

\pagebreak

\begin{table}
 \begin{center}\scriptsize
  \begin{tabular}{rrrr|rccccc|l}
   \hline\hline\noalign{\smallskip}
   \multicolumn{3}{c}{{Object Name}}&Detection Likelihood & \multicolumn{6}{c}{{X-ray Properties}} & {Class} \\
   \hline\noalign{\smallskip} {[PFH 2005]} & {[Stiele 2011]}&Chandra&LH& {f$_{\rm X}$} & {f$_{\rm X}$/f$_{\rm Opt}$} & {HR1} & {HR2} & {HR3} & {HR4} & \\
    &  &&  &{(mW/m$^2$)}
    & &  &  &  &  & \\
   \noalign{\smallskip} \hline\noalign{\smallskip}
 \textbf{620}&1598&     &25 & 8.4(1.0)e-15 &    & 0.90(0.22) & 0.19(0.18) & -0.23(0.18) &  0.05(0.33) & $<$hard$>$\\
&&&                         & 7.3(1.4)e-15 &    & 1.00(0.17) & 0.09(0.21) & -0.09(0.20) & -0.03(0.34) & $<$hard$>$\\
&&&                          &2.8(0.2) e-14 &   & 0.47(0.18)& 0.63(0.07)&-0.05(0.07)&-0.03(0.10) &$<$hard$>$ \\
 \textbf{705}&1744&     &58      & 1.2(0.2) e-14 &    &    -     & 0.74(0.14)  &-0.19(0.12)  & -0.30(0.25) & $<$hard$>$\\
&&&                              & 2.49(0.17)e-14&    & 0.71(0.31)&0.69(0.07) & -0.06(0.07)  & -0.39(0.09) & $<$hard$>$ \\
   \noalign{\smallskip} \hline\noalign{\smallskip}
   \noalign{\smallskip}
  \end{tabular}
 \end{center}
\end{table}

\begin{table}
 \begin{center}\scriptsize
 \caption{Optical properties of the stellar objects of Table \ref{Log}.}
  \label{Optical}
\scalebox{0.7}{
  \begin{tabular}{cccccccccclcccc}
   \hline\hline\noalign{\smallskip}
   \multicolumn{2}{c}{{Object Name}} & \multicolumn{9}{c}{{Optical Properties}}& \multicolumn{4}{c}{{Infrared Properties}}\\
   \hline\noalign{\smallskip} {oID} & {xID} &  &  &  &  &  &  &  & &&&&& \\
   {(LGGS)} & {(Piestch)} & {RA} & {Dec} & {V} & {B-V} & {U-B} & {V-R} & {R-I} & Q  &2 MASS&J&H&K\\
   \noalign{\smallskip} \hline\noalign{\smallskip}
LGGS\_J004123.75+411459.6  & \textbf{136}  & 00 41 23.75 & 41 14 59.6 & 20.293(0.011) & 1.862(0.031) &  0.761(0.081) & 1.469(0.012) &2.049(0.005)& -0.580 & 00412375+4114596&15.335(055)&14.865(076)&14.416(070)    \\
LGGS\_J004130.37+410500.9  & \textbf{146}  & 00 41 30.37 & 41 05 00.9 & 18.499(0.052) & 0.052(0.009) & -1.017(0.013) & 0.377(0.016) & -0.052(0.013)&-1.054  &&&&\\
LGGS\_J003937.29+404925.8 & \textbf{235-st}  & 00 39 37.3 & 40 49 25.8 & 20.468(0.012) & 0.064(0.016) & -0.876(0.012) & 0.098(0.018) & 0.086(0.014)  & -0.922   &&&&\\
LGGS\_J003937.27+404928.9 & \textbf{235-st}  & 00 39 37.3 & 40 49 28.9 & 20.952(0.020) & 0.207(0.024) & -0.688(0.018) & 0.198(0.027) & 0.271(0.018)  & -0.83   &&&&\\
LGGS\_J004557.04+414830.0 & \textbf{1716-st} & 00 45 57.04 & 41 48 30 & 20.017(0.009) &0.123(0.014) & -0.355(0.013) & 0.133(0.013) & 0.182(0.009)& -0.44 &  &&&& \\
LGGS\_J004210.24+405149.4 & \textbf{224a}  & 00 42 10.24 & 40 51 49.4 & 19.387(0.004) & 0.264(0.005) & -0.682(0.005) & 0.206(0.006) & 0.219(0.004)& -0.872 &&&&\\
LGGS\_J004210.38+405148.1 & \textbf{224b}  & 00 42 10.38 & 40 51 47.4 & 21.461(0.019)  &0.367(0.026) &0.199(0.026)  & 0.272(0.031)  & 0.291(0.023)&&&&&\\
LGGS\_J004216.78+404814.4 & \textbf{242}  & 00 42 16.78 & 40 48 14.4 & 19.419(004) &-0.104(005) & -0.993(004) &-0.013(006) &-0.010(005)& -0.918 & 00421645+4048113&16.381(117)&16.377:&15.151:\\
126022 & \textbf{246}  & 00 42 18.72 & 41 14  1.2 & 18.260(055) &-0.238(078) & -1.809(131) & 0.166(089) &      -     & -1.638 & 00421865+4114021&13.644(029)&13.158(034)&13.038(036)\\
LGGS\_J004231.23+410435.3 & \textbf{278}  & 00 42 31.23 & 41  4 35.3 & 19.935(0.005) & 0.296(0.007) & -0.632(0.006) & 0.299(0.008) & 0.530(0.007)& -0.845 & &&&\\
LGGS\_J004259.07+413731.4 & \textbf{367}  & 00 42 59.07 & 41 37 31.4 & 20.127(0.008) & 0.465(0.014) & -0.247(0.015) & 0.368(0.011) & 0.416(0.007)& -0.582  &00425913+4137321&15.945(087)&15.087(097)&14.709(104)\\
LGGS\_J004258.94+413727.3 &\textbf{367obj1}& 00 42 58.94& 41 37 27.3 & 19.854(0.011) & 0.757(0.021) & -0.647(0.025) & 1.217(0.012) & 1.401(0.005)& -1.192 & 00425892+4137273&15.505(0.0605)&14.641(0.071)&14.294(0.078)\\
LGGS\_J004309.86+411900.9  & \textbf{405}  & 00 43  9.86 & 41 19  0.9 & 18.098(0.014) & 1.034(0.017) &  0.819(0.017) & 0.731(0.017) & 0.644(0.010)&0.074 &00430986+4119006&15.431(074)&14.597(075)& 14.400(088)\\
LGGS\_J004310.50+413852.0 & \textbf{407}  & 00 43 10.50 & 41 38 52.0 & 19.957(0.008) & 0.467(0.015) & -0.703(0.015) & 0.268(0.012) &0.297(0.009) & -1.039  &&&&\\
LGGS\_J004333.64+411404.8 & \textbf{466}  & 00 43 33.64 & 41 14 04.8 & 19.495(004) & 0.544(006) & -0.296(006) & 0.337(006) & 0.323(005)& -0.688 & &&&\\
LGGS\_J004445.06+415153.4 & \textbf{581}  & 00 44 45.06 & 41 51 53.4 & 20.623(011) &-0.118(016) & -0.963(013) &-0.019(018) & 0.082(014)& -0.878 &&&&\\
LGGS\_J004456.78+413548.0 & \textbf{599}  & 00 44 56.78 & 41 35 48.0 & 20.534(011) & 0.073(019) & -0.893(018) & 0.196(017) & 0.555(011)& -0.946 & &&&\\
LGGS\_J004506.46+420615.8 & \textbf{612}  & 00 45 06.46 & 42 06 15.8 & 20.771(013) &-0.064(018) & -1.088(014) &-0.022(022) & 0.157(015)& -1.042 & &&&\\
LGGS\_J004513.59+413805.7 & \textbf{620}  & 00 45 13.59 & 41 38  5.7 & 19.932(0.007) & 0.164(0.011) &  0.121(0.012) & 0.140(0.010) & 0.242(0.007)   &  0.003 & &&&\\
LGGS\_J004604.83+415142.4 & \textbf{705}  & 00 46 04.83 & 41 51 42.4 & 18.145(0.004) & 0.481(0.004) & -0.062(0.004) & 0.328(0.004) & 0.294(0.004) & -0.408 &00460484+4151422&16.759(162)&16.855:&16.332:\\
\noalign{\smallskip} \hline\noalign{\smallskip}
\noalign{\smallskip}
  \end{tabular}
}
 \end{center}
\end{table}
\end{landscape}

\clearpage

\begin{table}
 \begin{center}
  \caption{Set of Spectral Standard Stars for Cross-Correlation}
  \label{standards}
  \begin{tabular}{cccc}
   \hline\hline\noalign{\smallskip}
Star             & RA (J2000)    & Dec. (J2000) & Spectral Type\\
\noalign{\smallskip}
\hline\noalign{\smallskip}
HD188001	 & 19:52.4 	 & +18:40 &  O7.5 Iaf \\
B+60 447   	 & 02:11.7 	 & +60:43 &  O9 Ia \\
B+61 154   	 & 00:43.3 	 & +61:55 &  O9.5 V \\
B+57 153   	 & 00:49.0 	 & +58:26 &  B0 V \\
B+63 33    	 & 00:21.8 	 & +64:36 &  B1 V \\
B+58 351  	 & 02:00.7 	 & +58:59 &  B1 III \\
B+60 191   	 & 01:14.5 	 & +61:20 &  B2 IV \\
VOR+59 97	 & 02:18.9 	 & +59:59 &  B2 V \\
VOR+59   	 & 02:16.8 	 & +59:46 &  B3 V \\
B+57 18  	 & 00:10.6 	 & +58:45 &  B3 IV \\
B+57 58  	 & 00:19.9 	 & +58:36 &  B5 V \\
L+60 221  	 & 02:10.5 	 & +61:07 &  B5 III \\
HD13799   	 & 02:16.8 	 & +62:57 &  B6 III\\ 
BSD 285   	 & 00:52.6 	 & +59:41 &  B7 V \\
B+59 6   	 & 00:10.9 	 & +60:26 &  B8 V \\
B+60 461   	 & 02:19.0 	 & +61:09 &  B8 IV \\
LATT.161	 & 00:44.7 	 & +61:54 &  B9 V \\
B+59 35   	 & 00:20.4 	 & +60:24 &  A0 V \\
B+60 51   	 & 00:26.6 	 & +61:25 &  A2 IB\\ 
VOR+61 44	 & 02:20.4 	 & +61:56 &  A3 V \\
REB 149   	 & 01:56.6 	 & +37:57 &   F5 V \\
BD+40349	 & 01:41.8 	 & +41:22 &  G0 V \\
\noalign{\smallskip} \hline\noalign{\smallskip}
\noalign{\smallskip}
  \end{tabular}
 \end{center}
\end{table}

\begin{table}
 \begin{center}
  \caption{Most significant lines identified in the optical spectrum of [PFH2005]~146.}
  \label{146_lines}
  \begin{tabular}{cccc}
   \hline\hline\noalign{\smallskip}
   \multicolumn{3}{c}{{Red channel}} \\
   \hline\noalign{\smallskip} Element & Wavelength & Flux & Vel.\\
   (\textbf{wavelength}) & {measured (\AA)} & (erg~s$^{-1}$) & km~s$^{-1}$ \\
   \noalign{\smallskip} \hline\noalign{\smallskip}
   $[$NII$]$\textbf{5755} & 5746.091 &  $(1.6\pm0.6)\times 10^{-15}$ & $-$470\\ %
   $[$NII$]$\textbf{6548} & 6538.275 & $(5.6\pm0.25)\times 10^{-14}$ & $-$460\\
   $[$NII$]$\textbf{6583} & 6573.52 &  $(1.6\pm0.1)\times 10^{-13}$ & $-$460\\
   $[$H${\alpha}]$\textbf{6563} & 6552.952 & $(3.5\pm0.2)\times 10^{-13}$ & $-$460\\
   $[$SII$]$\textbf{6716} & 6706.437 & $(5.6\pm0.2)\times 10^{-14}$& $-$450\\
   $[$SII$]$\textbf{6731} & 6720.711 & $(4.8\pm0.2)\times 10^{-14}$& $-$450\\
   \noalign{\smallskip} \hline\noalign{\smallskip}
   \multicolumn{3}{c}{{Blue channel}} \\
   \hline\noalign \\
   $[$H${\beta}]$\textbf{4861} & 4853.532 & $(7.2\pm0.1)\times 10^{-14}$& $-$460\\
   $[$OIII$]$\textbf{4959} & 4950.453 & $(6.8\pm0.4)\times 10^{-15}$& $-$510\\
   $[$OIII$]$\textbf{5007} & 4998.441 & $(1.8\pm0.1)\times 10^{-14}$& $-$510\\
    \noalign{\smallskip} \hline\noalign{\smallskip}
    \noalign{\smallskip}
  \end{tabular}
 \end{center}
\end{table}

\begin{table}
 \begin{center}
  \caption{Most significant Wolf-Rayet lines identified in the red optical spectrum of [PFH2005] 146.}
  \label{146_lines_wr}
  \begin{tabular}{cccc}
   \hline\hline\noalign{\smallskip}
   \multicolumn{3}{c}{{Red channel}} \\
   \hline\noalign{\smallskip} Element & Wavelength & Flux & Vel.\\
   (\textbf{wavelength}) & {measured (\AA)} & (erg~s$^{-1}$) & km s$^{-1}$\\
   \noalign{\smallskip} \hline\noalign{\smallskip}
   $[NII]$\textbf{5755} & 5746.091 & \nodata  & $-$470 \\
   $[NII]$\textbf{6548} & 6538. & $(1.4\pm0.1)\times 10^{-16}$ & $-$460\\
   $[NII]$\textbf{6583} & 6574 &  $(5.7\pm0.2)\times 10^{-16}$  & $-$410\\
   $[H{\alpha}]$\textbf{6563} & 6552.4 & $(5.0\pm0.4)\times 10^{-16}$ & $-$480 \\
   $[SII]$\textbf{6716} & 6706.3 & $(4.0\pm0.2)\times 10^{-16}$ & $-$450 \\
   $[SII]$\textbf{6731} & 6722.2 & $(1.4\pm0.4)\times 10^{-16}$ & $-$400\\
   \noalign{\smallskip} \hline\noalign{\smallskip}
   \multicolumn{3}{c}{{Blue channel}} \\
   \hline\noalign \\
   $[H{\beta}]$\textbf{4861} & - & - \\
   $[OIII]$\textbf{4959} & 4950.2 & $(1.9\pm0.3)\times 10^{-16}$ & $-$540\\
   $[OIII]$\textbf{5007} & 5000.3 & $(3.5\pm0.1)\times 10^{-16}$ & $-$420 \\
    \noalign{\smallskip} \hline\noalign{\smallskip}
    \noalign{\smallskip}
  \end{tabular}
 \end{center}
\end{table}

\begin{table}
 \begin{center}\scriptsize
  \caption{Most likely classification based on stellar spectra from Table \ref{Log}.}
  \label{Classifications}
  \begin{tabular}{ccccc}
   \hline\hline\noalign{\smallskip}
   Object Name & RA & DEC & Classification & Grade\\
   \hline
\textbf{136}  & 00 41 23.75 & 41 14 59.6 & Galactic Dwarf Nova & A\\
\textbf{146}  & 00 41 30.37 & 41 05 00.9 & WN+O colliding winds & B\\
\textbf{224}  & 00 42 10.24 & 40 51 49.4 & SNR & A\\
\textbf{235-st}  & 00 39 37.3 & 40 49 25.8 & M31 HMXB & A\\
\textbf{242}  & 00 42 16.78 & 40 48 14.4 & Galactic M star & B\\
\textbf{246}  & 00 42 18.72 & 41 14  1.2 & M31 LMXB in GC & A\\
\textbf{278}  & 00 42 31.23 & 41  4 35.3 & M31 HMXB & A\\
\textbf{367}  & 00 42 59.07 & 41 37 31.4 & M31 HMXB & B\\
\textbf{405}  & 00 43  9.86 & 41 19  0.9 & M31 symbiotic X-ray binary or LMXB & B\\
\textbf{407}  & 00 43 10.50 & 41 38 52.0 & M31 BeXRB & A\\
\textbf{466}  & 00 43 33.64 & 41 14 04.8 & M31 HMXB & B\\
\textbf{581}  & 00 44 45.06 & 41 51 53.4 & Galactic CV & A\\
\textbf{599}  & 00 44 56.78 & 41 35 48.0 & Bright source not counterpart & F\\
\textbf{612}  & 00 45 06.46 & 42 06 15.8 & M31 HMXB & B\\
\textbf{620}  & 00 45 13.59 & 41 38  5.7 &  M31 HMXB & B\\
\textbf{705}  & 00 46 04.83 & 41 51 42.4 & Bright source not counterpart & F\\
\textbf{1716-St} & 00 45 57.04 & 41 48  30 & M31 HMXB & B\\
\hline
\noalign{\smallskip}
  \end{tabular}
 \end{center}
\end{table}

\begin{table}
 \begin{center}\scriptsize
  \caption{Radial Velocities of the M31 members, along with the radial velocity of the gas disc at their projected locations from the 21 cm map of \citet{chemin2009}.}
  \label{velocities}
  \begin{tabular}{ccc}
   \hline\hline\noalign{\smallskip}
Object Name  & M31 H{\sc I} R.V. & R.V. (km s$^{-1}$)\\
   \hline
146 &  -460   &  -450\\
224 & -440  &   -450\\
235 & -450  &   -250\\
246 & -450  &   -350\\
278 & -450  &   -400\\
367 & -240  &   -350\\
405 & -120  &   -180\\
407 & -220  &   -140\\
466 & -290   &  -130\\
612 &  -110  &  -110\\
620 &  -130 &   -160\\
1716 & -120  &  -170\\
\hline
\noalign{\smallskip}
  \end{tabular}
 \end{center}
\end{table}

\clearpage

\clearpage

\clearpage

\clearpage

\end{document}